\documentclass{aa}  
\usepackage{graphicx}
\usepackage{float}
\usepackage{multicol,lipsum}
\usepackage{tabularx}
\usepackage{enumitem}
\usepackage{caption}
\usepackage{subcaption}
\usepackage{adjustbox}
\usepackage{makecell}
\usepackage{booktabs}
\usepackage{placeins}

\usepackage{txfonts}

\usepackage{url}
\usepackage[colorlinks,citecolor=blue,urlcolor=blue,bookmarks=false,hypertexnames=true]{hyperref}
\usepackage [english]{babel}
\usepackage [autostyle, english = american]{csquotes}
\MakeOuterQuote{"}

\begin{document} 

   \title{Dust characterization of halos: The extended emission in protoplanetary disks}

   \author{Sreejita Das \inst{1,2}
          \and
          Enrique Mac\'{\i}as \inst{1}
          \and
          Nicolas T. Kurtovic \inst{3}
          \and 
          Til Birnstiel \inst{2}
          \and 
          Elena M. Viscardi \inst{1}
          \and 
          Pietro Curone \inst{4}
          }

   \institute{European Southern Observatory, Karl-Schwarzschild-Strasse 2, 85748 Garching bei München, Germany\\
              \email{sreejita.das@eso.org}
         \and
             Ludwig-Maximilians-Universität München, Universitäts-Sternwarte, Scheinerstr. 1, 81679 München, Germany 
        \and
            Max Planck Institute for Astronomy, Königstuhl 17, 69117 Heidelberg, Germany
        \and 
            Departamento de Astronomía, Universidad de Chile, Camino El Observatorio 1515, Las Condes, Santiago, Chile
             }

  \abstract 
   {Extended low surface brightness emission has been identified in a number of protoplanetary disks, in tension with predictions of radial drift theory.}
   {We aim to investigate the nature and origin of faint, extended dust emission in the outer regions of protoplanetary disks, which we define as the ``Halo'', using multiwavelength (sub)millimeter continuum observations of three systems: Elias 2–24, IM Lup, and DM Tau.}
   {We utilized Atacama Large Millimeter Array (ALMA) observations of our targets to perform spectral energy distribution (SED) fitting with four dust compositions and derived radial profiles of their dust properties.}
   {The halos identified in our sources account for 20 - 30\% of  the total flux density at (sub)millimeter wavelengths. In Elias 2-24, IM Lup, and DM Tau, we infer maximum grain sizes of 2 cm, < 4 mm, and < 9 mm, with the data best reproduced by porous amorphous carbon, compact amorphous carbon, and compact organic carbon compositions, respectively. Their total dust masses are $125^{+34}_{-23}$, $301^{+139}_{-101}$, and $829^{+761}_{-378}$ M$_{\oplus}$, with corresponding halo masses of $33^{+12}_{-6}$, $103^{+25}_{-17}$, and $316^{+202}_{-117}$ M$_{\oplus}$. The halos of IM Lup and DM Tau are dust rich with gas-to-dust mass ratios of 64 and 18, respectively. In all three disks, the dust drift and growth timescales are shorter than the disk ages, implying that the smooth outer disks should not exist.}  
   {The halos in our sources hold relevant fractions of the total dust reservoir, demonstrating that they play an important role in alleviating the mass-budget problem. While the persistence of halos in IM Lup and DM Tau could be explained by late infall, the presence of centimeter-sized grains in Elias 2-24's halo suggests that unresolved dust traps also play a role.}


\maketitle

\section{Introduction}

Dust grains constitute the raw material from which planetesimals and, ultimately, planets form. In protoplanetary disks, repeated collisions and sticking allow solids to grow efficiently from micron sized particles to millimeter and centimeter scales over relatively short timescales \citep{brauer2008b, birnstiel2010, birnstiel2024, vorobyov2024}. The amount and spatial distribution of dust therefore set the initial conditions for planet formation. In particular, the total dust mass determines the reservoir of solids that can be converted into planetary cores, and it must be large enough to sustain an adequate flux of pebbles before the gaseous disk disperses \citep{lambrechts2014, xu2017, shibata2025}.

While the total dust mass is a key quantity for planet formation, it can only be robustly inferred if the full radial extent of the dust distribution is traced. Disk sizes measured with the Atacama Large Millimeter/sub-Millimeter Array (ALMA) are often used as a proxy for the spatial distribution of solids (see, e.g., \citealp{hendler2020} and references therein). However, these sizes are strongly influenced by observational sensitivity rather than marking a physical outer edge of the disk (e.g., the case of TW Hya; \citealp{das2024}). For example, \citealp{rosotti2019} showed that commonly reported disk radii trace the region where grains remain large enough to have a significant submillimeter opacity, rather than the true extent of the dust surface density. If true, sufficiently sensitive observations should reveal low surface brightness emission in the outer disk, potentially revising current estimates of disk sizes and dust masses.

Understanding how disks evolve and attain large radial extents is essential to interpreting faint emission in their outer regions. In the classical viscous spreading theory, angular momentum transport, parameterized through an effective viscosity, drives inward mass accretion while simultaneously forcing a fraction of the disk material to move outward to conserve angular momentum, leading to an overall increase in disk size over time (\citealp{shakura1973, pringle1974}; \citealp{manara2023} for a review). An alternative picture invokes magnetohydrodynamic (MHD) disk winds, which remove mass from the disk surface and extract angular momentum vertically rather than transporting it radially, potentially limiting disk size growth while still allowing efficient stellar accretion (e.g., \citealp{turner2014}; \citealp{pascucci2023} and references therein). 

In addition to the disk evolution theories discussed above, extended outer disks may arise from the late infall of material from the surrounding environment, as indicated by recent observations (e.g., \citealp{gupta2023, winter2024}). Due to the high angular momentum of the remnant envelope captured by the star, this material does not fall directly onto the star but instead orbits around it, potentially forming a secondary dust disk surrounding the already existing primordial disk \citep{kuffmeier2020}. In this scenario, the outer disk is replenished with small grains at late times, producing low surface brightness emission without requiring long-term viscous expansion. Such replenishment could explain the dichotomy of meteorites (e.g., \citealp{nanne2019}), the misalignment of some disks and planetary systems (e.g., \citealp{kuffmeier2021}), and, most importantly, alleviate the mass-budget problem \citep{manara2018, mulders2021}. If correct, extended dust disks do not simply trace the evolutionary outcome of primordial material but instead record ongoing interaction between disks and their environment. 

An informative approach to understanding the origin and persistence of extended dust emission involves comparing the dust drift and growth timescales. If extended emission arises from long-term viscous spreading of primordial material, we expect long drift and growth timescales, or equivalently efficient trapping mechanisms such as pressure bumps to prevent the rapid depletion of solids at large radii \citep{pinilla2012, rosotti2020, pinilla2025}. In contrast, if the outer disk is replenished through late infall, the observed dust may not yet be in steady state, allowing extended emission to persist even when local drift timescales are shorter than the disk age. The grain size distribution offers an additional diagnostic as drift-dominated evolution is expected to produce a characteristic size distribution slope of 2.5, while fragmentation-dominated regimes can yield steeper slopes of 3.5 \citep{birnstiel2012}.

In this paper, we use multiwavelength (sub)millimeter continuum observations of three protoplanetary disks - Elias 2–24, IM Lup, and DM Tau - to investigate the nature and origin of faint extended dust emission in their outer regions, which we refer to as the ``Halo''. These halos in Class II disks are different from debris disk halos, which are generally expected to consist of small grains generated by collisions within a planetesimal belt and dispersed onto highly eccentric orbits by radiation pressure (e.g., \citealp{thebault2023}). We performed spectral energy distribution (SED) fitting to constrain the dust properties, combined these results with gas surface density profiles from the literature to infer gas-to-dust ratios, and computed drift and growth timescales to assess the physical origin of the halo emission. 

The paper is structured as follows: In Sect. \ref{2}, we describe the observations used in this work, along with the calibration and imaging techniques. In Sect. \ref{3}, we analyze the dust continuum with brightness temperature profiles and spectral index radial profiles. In Sect. 4, we outline the method used for dust characterization and follow it with our results. In Sect. \ref{5}, we discuss the implications of our results, and finally we summarize our conclusions in Sect. \ref{6}. 

\begin{figure*} 
  
    \includegraphics[width=1.0\textwidth]{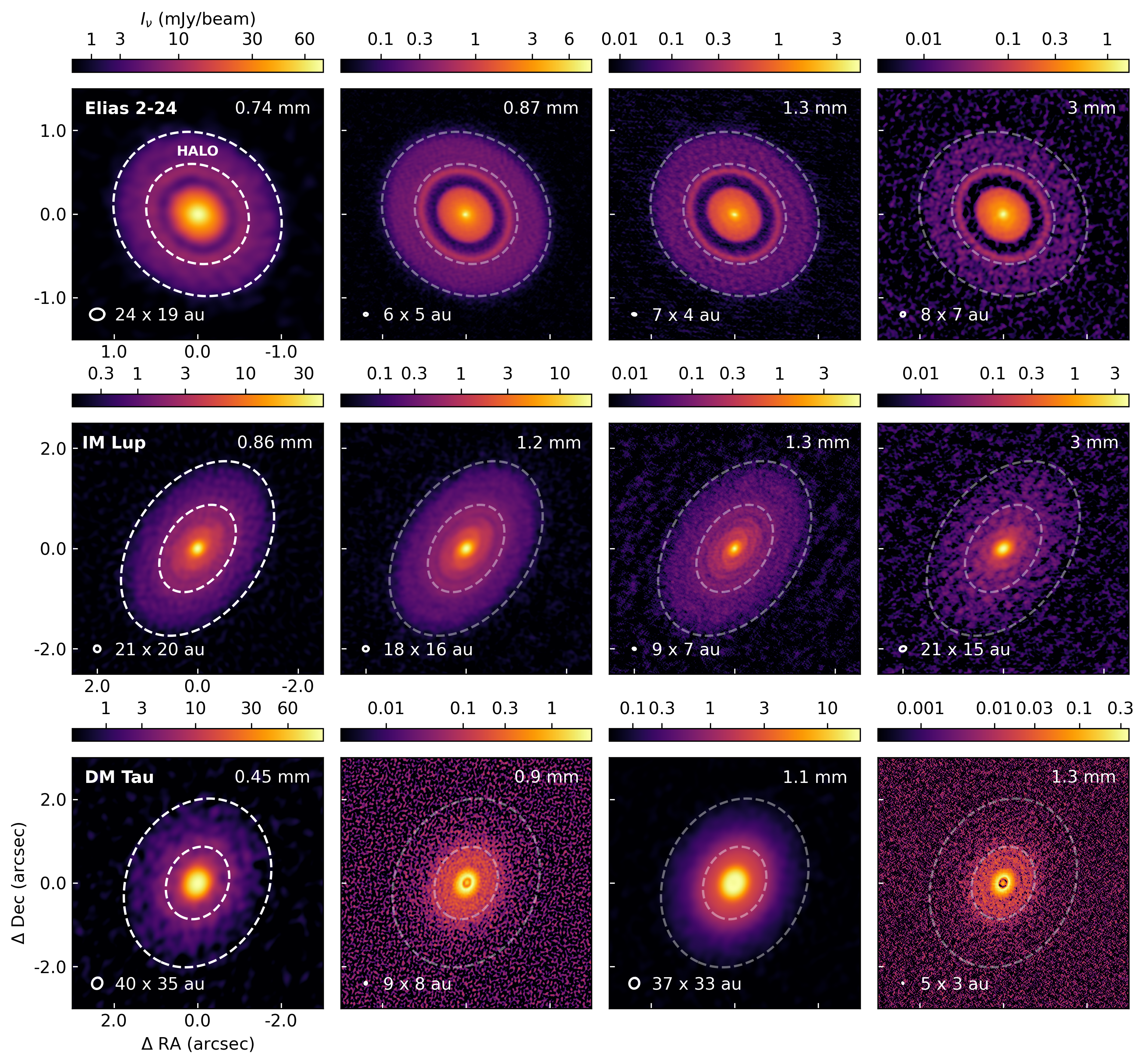} 
    
    \caption {{Gallery of the continuum images, obtained with the CLEAN algorithm at the highest resolution possible. The properties of these images are listed in Table \ref{table:images}. Dashed ellipses mark the extent of the halo (see Sect. \ref{3} for its definition). The beam dimensions are shown in the bottom-left corner of each plot. We note that logstrech was used to enhance the faint extended emission.}} \label{fig:gallery}
\end{figure*}

\section{Observations} \label{2} 
\subsection{Sample selection}
We used 18 archival dust continuum observations of three protoplanetary disks: Elias 2-24, IM Lup, and DM Tau. Beyond their outermost substructures, these disks exhibit smoothly declining intensity profiles with no obvious ring-like features, yet they have detectable extended continuum emission. We ruled out contamination of the extended emission by envelope emission based on its axisymmetry, vertically thin geometry, and Keplerian rotation of the $^{12}$CO gas disk well beyond the continuum emission (Elias 2-24: \citealp{pinte2023}; IM Lup: \citealp{martire2024}; DM Tau: \citealp{longarini2025}). For the dust characterization, it was essential that each source had archival data in at least three ALMA bands. Given enough sensitivity, when the datasets had a frequency separation of $>$ 20 GHz, they were treated as independent measurements. In addition, we only included observations with angular resolutions < 0.3'', which ensured that the compact inner disk features could be spatially separated from the faint extended emission. {All observations contained short baseline data to be sensitive to larger spatial scales.} Details about the observations used in this work can be found in Table \ref{table:data}.

\subsection{Calibration}
To improve the quality of the ALMA data, we applied self-calibration with the Common Astronomy Software Applications v6.5.4 (\texttt{CASA}; \citealp{McMullin2007}). The continuum was extracted by flagging channels within $\pm$ 25 km s$^{-1}$ around the targeted emission lines, and the data were subsequently averaged over 125 MHz channel widths, following the same procedure as \citealp{andrews2018}. The source coordinates were determined by querying the GAIA DR3 archive \citep{gaiacollab2018} via the \texttt{astroquery} package. These coordinates were used to align the phase center of each observation with the stellar position using the \texttt{phaseshift} task. For consistency, the phasecenters of all observations of Elias 2-24, IM Lup, and DM Tau were corrected to J2000 16h26m24.07s -24d16m13.50s, J2000 15h56m09.19s -37d56m06.53s, and J2000 04h33m48.75s +18d10m09.56s, respectively, following \citealp{dsharp2} and \citealp{curone2025}, using the \texttt{fixplanets} task. When the short baseline observations showed a flux variation greater than 10\% relative to the long-baseline data, they were self-calibrated independently before being combined with the long-baseline data. After combining the short and long baseline observations, self-calibration was applied again. All spectral windows and scans were combined to maximize the S/N. Several rounds of phase self-calibration were performed, starting with a solution interval of infinite and progressively decreasing (to multiples of the integration time and finally to \texttt{solint=‘int’}) until no further improvement in S/N was achieved. After each step, the image was cleaned down to 1$\sigma$, and the resulting model was saved to the model column to compute the next calibration table. A final round of amplitude self-calibration with an infinite solution interval was also applied. 

Some of the observations taken directly from previously published works are: Elias 2-24 from \citealp{andrews2018} (2016.1.00484.L) and \citealp{carvalho2024} (2017.1.01330.S, 2018.1.01198.S); IM Lup from \citealp{oberg2021} (2018.1.01055.L), \citealp{andrews2018} (2016.1.00484.L), and \citealp{careno2024} (2019.1.01357.S); and DM Tau from \citealp{teague2025} (2021.1.01123.L). Furthermore, the Band 6 and 9 projects of DM Tau were self-calibrated individually with the ALMA pipeline in CASA v6.6.1 due to their high S/N and then self-calibrated again by the method above after concatenation. Their phasecenters were also corrected in the same way as described above. Observations with a central frequency separation of less than 20 GHz were merged, and datasets that were too large for efficient processing were further reduced using the \texttt{split} task.

\begin{table*}
\caption{Imaging results.}
\label{table:images}
\centering
\begin{tabular}{c c c c c c c c}     
\hline\hline       
                     
\thead{Band} & \thead{Wavelength \\ (mm)\tablefootmark{a}} & \thead{robust} &  \thead{Beam dimensions \\ (mas, au (\textdegree))} & \thead{Peak $I_{\nu}$ \\ (mJy beam$^{-1}$), K\tablefootmark{b}} & \thead{rms \\ ($\mu$Jy beam$^{-1}$), K\tablefootmark{b}} & \thead{Peak \\ S/N} & \thead{Flux density \\ (mJy)\tablefootmark{c}} \\ [3ex]

\midrule
\multicolumn{8}{c}{\textbf{Elias 2-24}} \\ 
\midrule
8 & 0.740 & 0.3 & 171 $\times$ 137, 21 $\times$ 19 (-83.7) & 78.3, 33.7 & 26.3, 2.5 & 298 & 1014 $\pm$ 0.2 ($\pm$ 203) \\ 
7 & 0.868 & 0.5 & 44 $\times$ 34, 6 $\times$ 5 (-88.6) & 9.2, 69.6 & 11.1, 3.1 & 824 & 794 $\pm$ 0.4 ($\pm$ 79) \\ 
6 & 1.27 & 0.5 & 47 $\times$ 27, 7 $\times$ 4 (78.1) & 4.7, 85.7 & 16.4, 3.1 & 284 & 401 $\pm$ 0.6 ($\pm$ 40) \\ 
3 & 3.04 & 0.5 & 56 $\times$ 52, 8 $\times$ 7 (-44.3) & 1.6, 72.5 & 7.2, 1.7 & 226 & 37 $\pm$ 0.2 ($\pm$ 2) \\ 

\midrule
\multicolumn{8}{c}{\textbf{IM Lup}} \\ 
\midrule
7 & 0.855 & -0.6 & 132 $\times$ 128, 21 $\times$ 20 (33.9) & 42.9, 33.0 & 67.0, 2.8 & 641 & 605 $\pm$ 0.9 ($\pm$ 60) \\ 
6 & 1.17 & 0.5 & 114 $\times$ 104, 18 $\times$ 16 (-86.0) & 20.7, 38.3 & 15.4, 2.0 & 1349 & 255 $\pm$ 0.2 ($\pm$ 26) \\ 
6 & 1.28 & 0.5 & 68 $\times$ 54, 9 $\times$ 7 (82.0) & 10.1, 66.5 & 10.0, 2.1 & 1011 & 243 $\pm$ 0.3 ($\pm$ 24) \\ 
3 & 3.06 & 0.3 & 132 $\times$ 98, 21 $\times$ 15 (-69.4) & 4.4, 45.5 & 78.2, 2.4 & 564 & 21 $\pm$ 1.2 ($\pm$ 1) \\

\midrule
\multicolumn{8}{c}{\textbf{DM Tau}} \\ 
\midrule
9 & 0.446 & 0.5 & 275 $\times$ 241, 40 $\times$ 35 (-24.0) & 115.6, 15.7 & 432.0, 4.3 & 267 & 664 $\pm$ 4.0 ($\pm$ 133) \\ 
7 & 0.898 & -0.5 & 63 $\times$ 55, 9 $\times$ 8 (-15.7) & 2.8, 15.5 & 40.3, 3.3 & 69 & 220 $\pm$ 2.0 ($\pm$ 22) \\ 
7 & 1.06 & 0.5 & 257 $\times$ 233, 37 $\times$ 33 (-16.8) & 18.5, 10.0 & 13.1, 1.6 & 1411 & 169 $\pm$ 0.1 ($\pm$ 17) \\ 
6 & 1.34 & 0.5 & 33 $\times$ 22, 5 $\times$ 3 (30.4) & 0.4, 18.3 & 7.1, 2.8 & 53 & 103 $\pm$ 0.6 ($\pm$ 10) \\

\hline\hline    
\vspace{0.1mm}
\end{tabular}
\newline Notes: \tablefoottext{a}{The images are shown in Fig. \ref{fig:gallery}.} \tablefoottext{b}{The peak and rms brightness temperatures are computed using the full Planck function (Eq. \ref{eqn:planck}). The frequencies used in this calculation are obtained with the analysisUtils function \texttt{getWeightedMeanScienceFrequency()} applied to the respective ms files.} \tablefoottext{c}{The first set of uncertainties is calculated as the rms multiplied by the square root of the number of synthesized beams within the mask. The second set of uncertainties, shown in parentheses, additionally accounts for the absolute flux calibration uncertainties given in the ALMA Proposer’s Guide: 20\% for Bands 9 and 8, 10\% for Bands 7 and 6, and 5\% for Band 3.}
\end{table*}

\begin{figure*} 
    
    \includegraphics[width=1.0\textwidth]{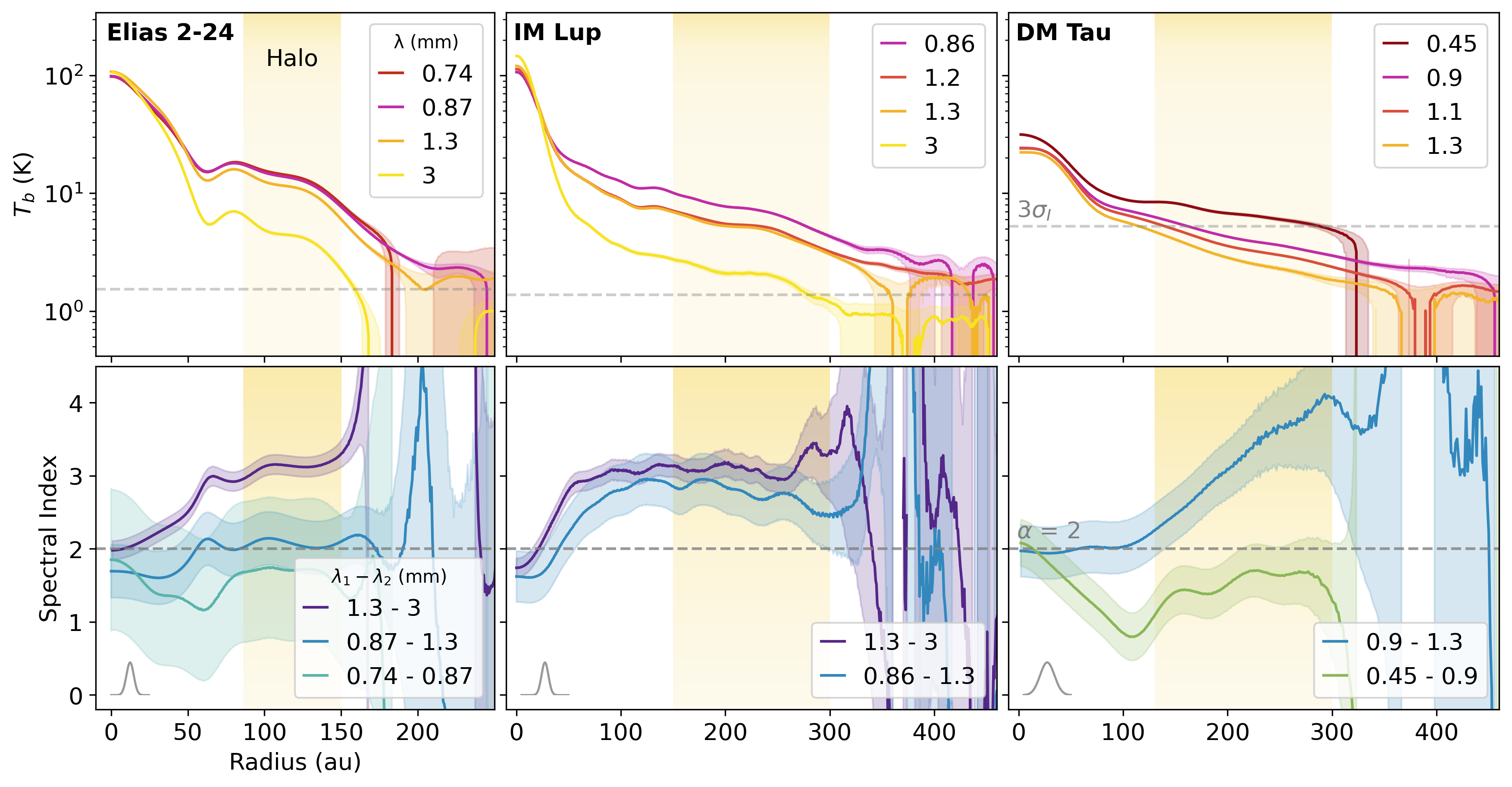} 
    
    \caption {\textit{Top row:} Brightness temperature, $T_\mathrm{b}$, profiles of our sample at all wavelengths listed in Table \ref{table:images} calculated using the full Planck function in Eq. \ref{eqn:planck}. For each source, the images were smoothed to the lowest angular resolution. The shaded regions represent the uncertainties of the individual profiles. Absolute flux calibration uncertainties are not included here. The yellow shaded bands mark the halo defined for each source. The radial axis extends to the largest radius at which emission is detected in the most sensitive observation: 240 au at 0.87 mm, 450 au at 0.86 mm, and 460 au at 0.9 mm in Elias 2-24, IM Lup, and DM Tau, respectively. The dashed gray line marks the 3$\sigma_\mathrm{I}$ of the least sensitive observation (3 mm for Elias 2-24, 3 mm for IM Lup, and 0.45 mm for DM Tau), where $\sigma_\mathrm{I}$ is the rms of the respective intensity profile. \textit{Bottom row:} Spectral index profiles computed from contiguous pairs of the intensity profiles shown in the top row. The shaded regions indicate the uncertainties obtained through error propagation of the intensity profiles, accounting for both the rms and the absolute flux calibration uncertainties. $\alpha$ = 2 is marked with a gray dashed line. The bottom-left corner shows the gaussian profile of the beam, in gray, used to smooth their parent images.} \label{fig:profiles}
\end{figure*} 

\begin{table}
\caption{Disk sizes.}
\label{table:disk_sizes}
\centering
\begin{tabular}{c c c c c}     
\hline\hline       
                      
\thead{Disk} & \thead{Wavelength \\ (mm)\tablefootmark{a}} & \thead{$R_{68}$ (au)} &  \thead{$R_{90}$ (au)} & \thead{$R_{99}$ (au)}\\ [3ex]

\midrule
\vspace{1mm}
Elias 2-24 & 0.87 & 98.4$^{+0.2}_{-0.2}$ & 133.1$^{+0.3}_{-0.3}$ & 164.8$^{+2.0}_{-1.8}$ \\ 
\vspace{1mm}
IM Lup & 1.2 & 160.3$^{+1.9}_{-1.9}$ & 246.9$^{+2.6}_{-2.6}$ & 347.7$^{+11.2}_{-13.4}$ \\ 
\vspace{1mm}
DM Tau & 0.9 & 123.0$^{+2.3}_{-2.3}$ & 212.6$^{+6.9}_{-6.8}$ & 346.6$^{+30.3}_{-30.5}$ \\ 

\hline\hline    
\vspace{0.1mm}
\end{tabular}
\newline Notes: \tablefoottext{a}{We calculate the disk sizes for the observation with the highest sensitivity.}
\end{table}

\subsection{Imaging} \label{2.3} 

The final imaging was performed with the \texttt{tclean} task in CASA v6.6.5. When the spectral windows were separated by more than 10 GHz in frequency, the \texttt{mtmfs} deconvolver \citep{Rau2011} with \texttt{nterms} = 2 was used, assuming that the frequency dependence of the emission can be represented by a second-order Taylor expansion. For datasets where the spectral windows were closer in frequency, the \texttt{multiscale} deconvolver was adopted. For each observation, several Briggs weighting values were tested to achieve an optimal balance between sensitivity and angular resolution. For all sources, the pixel size was set to one-tenth of the synthesized beam of the highest-resolution dataset, and the image size was fixed to 10''. Multiscale clean components with scales of zero, two, six, and 12 times the beam size were used, with a slight bias toward smaller scales to better recover compact emission. The final values of the \texttt{robust} parameter, the resulting synthesized beam sizes, rms values, peak S/N, and integrated flux densities for each observation are listed in Table \ref{table:images}, and the resulting images are shown in Fig. \ref{fig:gallery}. For the purposes of dust characterization, various uv-taper values were explored to smooth the high resolution observations to match the lowest resolution data. The final common beam size was achieved with the \texttt{imsmooth} task on the uv-tapered images, ensuring that all images of the same source have the same angular resolution. 

\begin{table}
\caption{Flux density of the halo.}
\label{table:halo_flux}
\centering
\begin{tabular}{c c c}     
\hline\hline       
                      
\thead{Disk name} & \thead{Wavelength \\ (mm)} & \thead{Flux density \\ of halo (mJy)\tablefootmark{a}} \\ [3ex]

\midrule
Elias 2-24 & 0.74 & 371.4 $\pm$ 1.3 ($\pm$ 74.3) \\ 
& 0.87 & 288.8 $\pm$ 0.1 ($\pm$ 28.9) \\
& 1.3 & 128.7 $\pm$ 0.2 ($\pm$ 12.9) \\
& 3 & 8.4 $\pm$ 0.1 ($\pm$ 0.4) \\
\midrule
IM Lup & 0.86 & 220.1 $\pm$ 0.8 ($\pm$ 22.0) \\ 
& 1.2 & 83.5 $\pm$ 0.2 ($\pm$ 8.3) \\
& 1.3 & 70.7 $\pm$ 0.2 ($\pm$ 7.1) \\
& 3 & 4.8 $\pm$ 0.1 ($\pm$ 0.2) \\
\midrule
DM Tau & 0.45 & 186.1 $\pm$ 3.0 ($\pm$ 37.2) \\ 
& 0.9 & 66.5 $\pm$ 0.2 ($\pm$ 6.7) \\
& 1.1 & 40.5 $\pm$ 0.1 ($\pm$ 4.1) \\
& 1.3 & 21.0 $\pm$ 0.2 ($\pm$ 2.1) \\

\hline\hline    
\vspace{0.1mm}
\end{tabular}
\newline Notes: \tablefoottext{a}{We calculated the flux densities by placing masks at the inner and outer edges of the halo using the CASA task \texttt{imstat}. To make them comparable, we assume the same radial extent at each wavelength: 86 to 150 au for Elias 2-24, 150 to 300 au for IM Lup, and 130 to 300 au for DM Tau (see Sect. \ref{3} for the definition of the halo boundaries). The uncertainties are calculated the same way as Table \ref{table:images}}. 
\end{table}

\section{Radial profiles} \label{3} 

After imaging, we generate azimuthally averaged intensity profiles by deprojecting the images using the disk's inclination and position angle derived from literature. Specifically, we adopted an inclination angle and position angle of 29.0\textdegree \space and 45.7\textdegree \space for Elias 2-24, 47.5\textdegree \space and 144.5\textdegree \space for IM Lup \citep{dsharp2}, and 36.0\textdegree \space and 155.6\textdegree \space for DM Tau \citep{curone2025}, respectively. The uncertainty at each radius is given by $rms/(2\sqrt{n})$ where $rms$ is the image noise and $n$ denotes the number of beams within the annulus over which the intensities are averaged to construct the radial profile. For each source, we define the starting radius of the halo qualitatively as the region in the outer disk beyond the last ring identified in the high resolution intensity profile, where dust evolution is not dominated by observable pressure bumps and the emission transitions into a smooth, featureless profile. {We note that our definition is designed to describe the extended emission in this paper. Recent studies suggest that substructures in smooth disks may remain undetected due to limited resolution \citep{alvarado2025, quincy2026} or high optical depth (e.g., MP Mus; \citealp{ribas2025}), in which case defining a halo would be ambiguous.} The outer boundary of the halo is defined as the radius at which the azimuthally averaged intensity profile of the least sensitive observation drops to 3$\sigma_\mathrm{I}$, where $\sigma_\mathrm{I}$ is the rms of the respective intensity profile. {We note that the outer boundary is defined for the SED analysis in Sect. \ref{4} and does not hold any physical significance. With greater sensitivity, the emission would likely extend farther.} The halo emission is about 100 times fainter than the peak intensity. In terms of flux density, the halos in our sources account for 20 - 30\% of the total disk flux at ALMA wavelengths. The disk sizes $R_{68}$, $R_{90}$, and $R_{99}$, estimated as the radius enclosing 68\%, 90\%, and 99\% of the total flux density, are computed for the most sensitive observation and are listed in Table \ref{table:disk_sizes}. Additionally, the flux densities of the halo, measured assuming the same radial extent at each wavelength, are listed in Table \ref{table:halo_flux}.  

To enable a consistent comparison across wavelengths, we also generate intensity profiles from the smoothed images. From these profiles, we derive the brightness temperature profiles using the full Planck function, since we employ 0.45 mm data (ALMA Band 9) where the assumptions of low temperature and low frequency inherent to the Rayleigh–Jeans approximation no longer hold. The brightness temperature, $T_\mathrm{b}$, was computed as:  

\begin{equation}\label{eqn:planck}
T_\mathrm{b} = \frac{h \nu}{k_{\mathrm{B}}} 
\left[ 
\ln{\left( 1 + \frac{2 h \nu^{3}}{I_{\nu} c^{2}} \right)} 
\right]^{-1},
\end{equation} \\

where $h$ is the Planck constant, $k_{\mathrm{B}}$ is the Boltzmann constant, $\nu$ is the observing frequency, $c$ is the speed of light, and $I_{\nu}$ is the specific intensity expressed in $\mathrm{Wm^{-2}Hz^{-1}sr^{-1}}$. These profiles are plotted in the top row of Fig. \ref{fig:profiles}. The longer wavelength observations show a higher brightness temperature in the inner regions of the disk where it might be dominated by free-free emission as seen at r < 50 au in Elias 2-24 and r < 20 au in IM Lup (similar free–free contamination at longer wavelengths has been reported in TW Hya; \citealp{macias2021}, and HL Tau; \citealp{alvarado2024}). Furthermore, smoothing the images introduces beam dilution, which can suppress the brightness temperature. This effect is an inherent limitation of our modeling approach. In addition, we derive spectral index maps from the smoothed profiles as shown in the bottom row of Fig. \ref{fig:profiles}. 

\paragraph{Elias 2-24.} We recovered the gap-ring pair D58 and B78, respectively, that was previously identified in \citealp{dsharp2} (distance = 139 pc; \citealp{gaiadr3}). We identified the halo as the outermost regions of the disk starting at 86 au. In the most sensitive data at 0.87 mm, the disk emission extends up to 240 au. The 1.3 - 3 mm spectral index profiles show similar features as those identified in \citealp{carvalho2024}: 2 < $\alpha$ < 3 at r < 50 au with an increasing trend, a local maxima at D58, a local minima at B78 and a flat $\alpha$ $\sim$ 3 at the halo. In sharp contrast, the 0.87 - 1.3 mm and 0.74 - 0.87 mm profiles show very optically thick emission with $\alpha$ < 2 at r < 50 au and a flat $\alpha$ $\sim$ 2 at r > 50 au. 

\paragraph{IM Lup.} We recovered the spiral along with the gap-ring pair D114 and B132, respectively (distance = 155.8 pc; \citealp{gaiadr3}). These substructures were also previously identified in \citealp{dsharp2}. We consider the outermost region of the disk, starting at 150 au, as the halo. In the halo, the spirals are no longer visible. In the most sensitive 0.86 mm observations, the disk emission extends substantially farther, reaching out to 450 au. The 1.3 – 3 mm and 0.86 – 1.3 mm spectral index profiles exhibit similar trends: an overall increase in $\alpha$ from approximately 2 to 3 at r < 100 au, followed by a flat region with $\alpha$ $\sim$ 3 at r > 100 au. Within the uncertainties, the 0.86 – 1.3 mm profile also shows a slight decrease in $\alpha$ at 250 < r < 300 au, with values ranging between 2 and 3.  

\paragraph{DM Tau.} We recovered the gap–ring pairs D14 – B24, D72 – B90, and D102 – B11, which were also reported in \citealp{jun2021}, \citealp{francis2022}, and \citealp{curone2025}. Interestingly, despite the moderate resolution of our 0.45 mm data, we identify an additional gap and ring D119 and B140, respectively (distance = 144 pc; \citealp{gaiadr3}). These substructures are not evident in the 0.9 mm and 1.3 mm profiles, despite their higher resolution, indicating that the 0.45 mm observations might trace a different vertical layer of the disk due to its higher optical depth (see Fig. \ref{fig:optical_depth} in Appendix \ref{appendix_opacity} where $\tau \sim 0.3$ near B140). The continuum images and intensity profiles highlighting these substructures are provided in Appendix \ref{appendix_dmtau}. The outermost region of the disk starting at 130 au is defined as the halo. As shown in \citealp{curone2025}, continuum emission at 0.9 mm extends out to 460 au. The 0.9 – 1.3 mm spectral index profile shows a flat $\alpha \sim$ 2 at r < 100 au, followed by an overall increase in $\alpha$ from 2 to 4 at 100 < r < 300 au. In contrast, the 0.45 – 0.9 mm spectral index profile displays a behavior that can only be explained with scattering \citep{sierra2020}: a decreasing $\alpha$ from 2 to 0.8 at r < 120 au; a subsequent increase in $\alpha$ from 0.8 to 1.5 at 120 < r < 200 au; and, at 200 < r < 300 au, a nearly flat $\alpha \sim$ 1.7. 

\section{Dust characterization} \label{4} 

Multiwavelength continuum observations provide an effective way to constrain the following dust properties in protoplanetary disks: the dust temperature $T_\mathrm{d}$, the dust surface density $\Sigma_\mathrm{d}$, the maximum grain size $a_\mathrm{max}$, and the slope of the particle size distribution $q$ (e.g., \citealp{macias2021}; \citealp{guidi2022}; \citealp{sierra2024}; \citealp{viscardi2025}; \citealp{zagaria2025}; \citealp{shi2026}). In this work, we model the azimuthally averaged radial intensity profiles at each wavelength by assuming an axisymmetric, geometrically thin, and vertically isothermal disk. These assumptions are well justified at (sub)millimeter wavelengths, where the continuum emission primarily traces the midplane layer in which the largest dust grains have settled \citep{pinte2016}\footnote{We note that the 0.45 mm observation used for DM Tau may probe a different vertical layer of the disk. However, following \citealp{viscardi2025}, we expect the dust properties to remain robustly constrained, as the analysis also includes the more optically thin 1.3 mm observations (see Fig. \ref{fig:optical_depth}).}. Under these conditions, the observed intensity can be described by a one–dimensional, vertically isothermal slab model \citep{sierra2020}. The emergent intensity at frequency $\nu$ can thus be computed as a function of the above four parameters as follows: 

\begin{equation}
I_{\nu} = B_{\nu}(T_\mathrm{d}) \left[ (1 - e^{-\tau_{\nu}/\mu}) + \omega_{\nu} F(\tau_{\nu}, \omega_{\nu}) \right],
\end{equation}

where $B_{\nu}(T_\mathrm{d})$ is the Planck function, and $\tau_{\nu} = \Sigma_\mathrm{d} \chi_{\nu}$ is the optical depth. Here $\chi_{\nu}$ is the total dust opacity, defined as the sum of the absorption and scattering opacities ($\chi_{\nu} = \kappa_{\nu} + \sigma_{\nu}$). The dust albedo is given by $\omega_{\nu} = \sigma_{\nu}/(\kappa_{\nu} + \sigma_{\nu})$, and $\mu = \cos(i)$, where $i$ denotes the inclination angle ($i = 0^{\circ}$ for a face–on disk). The scattering contribution is described by the function $F(\tau_{\nu}, \omega_{\nu})$, expressed as:

\begin{equation}
\begin{split}
F(\tau_{\nu}, \omega_{\nu}) 
= \frac{1}{
\exp\!\left(-\sqrt{3}\,\epsilon_{\nu}\,\tau_{\nu}\right)(\epsilon_{\nu} - 1) - (\epsilon_{\nu} + 1)
} \space \times
\\[6pt] 
\Bigg[
\frac{
1 - \exp\!\left(-(\sqrt{3}\,\epsilon_{\nu} + 1/\mu)\tau_{\nu}\right)
}{
\sqrt{3}\,\epsilon_{\nu}\mu + 1
}
+
\frac{
\exp\!\left(-\tau_{\nu}/\mu\right) - \exp\!\left(-\sqrt{3}\,\epsilon_{\nu}\tau_{\nu}\right)
}{
\sqrt{3}\,\epsilon_{\nu}\mu - 1
}
\Bigg],
\end{split}
\end{equation}

where $\epsilon_{\nu} = \sqrt{1 - \omega_{\nu}}$. To account for anisotropic scattering, we adopt the effective scattering coefficient $\sigma_{\nu}^{\mathrm{eff}} = (1 - g_{\nu})\,\sigma_{\nu}$, where $g_{\nu}$ is the scattering asymmetry parameter that quantifies the degree of forward scattering \citep{carrasco2019}. 

To link this formalism with the observed multiwavelength emission, we compute the dust opacity laws based on a prescribed dust composition and grain size distribution (\citealp{beckwith1990}). For a given composition and assuming a power–law grain size distribution $n(a) \propto a^{-q}$, where $a$ is the particle radius, the absorption and scattering opacities are calculated as functions of the maximum grain size $a_\mathrm{max}$ and the power law slope $q$. 

In this work, we fit our free parameters using four dust compositions: DSHARP with no porosity, DSHARP with 70\% porosity, ZUBKO with no porosity, and ZUBKO with 70\% porosity. In the following sections, these compositions are referred to as DSHARP (compact), DSHARP (70\% porosity), ZUBKO (compact), and ZUBKO (70\% porosity), respectively. The DSHARP grains are composed of 36\% water ice \citep{warren2008}, 16\% silicates \citep{draine2003}, 2\% troilite, and 44\% refractory organic carbon \citep{henning1996} by volume fraction \citep{birnstiel2018}. For the ZUBKO grains, we adopt the same volume fractions and bulk densities as in \citealp{birnstiel2018}, but replace the organic carbon component with amorphous carbon (\citealp{zubko1996}, BE sample) and assume an internal grain density of 2.5 g cm$^{-3}$ \citep{ricci2010}. For the porous compositions, we compute the opacities in two steps: first mixing the solid constituents using the \textit{Bruggeman} rule, and then introducing porosity via the \textit{Maxwell–Garnett} rule \citep{bohren1998}. All opacity calculations are performed using the \texttt{get\_opacities} function in the \texttt{dsharp\_opac} Python package. 

The intensity at each radius (from Sect. \ref{3}) is fitted independently, without assuming any specific radial profile for temperature or surface density. Thus, to ensure physically plausible temperatures that are consistent with stellar irradiation, we adopted a prior on the dust temperature $T_\mathrm{d}$ based on the temperature profile of a passively irradiated flared disk that is in radiative equilibrium \citep{dullemond2001}:

\begin{equation} \label{eqn:prior}
T_\mathrm{d}(r) = 
\left(
\frac{\phi L_\ast}{8\pi r^2 \sigma_\mathrm{SB}}
\right)^{0.25},
\end{equation}

where $\phi$ is the disk flaring angle, $L_\ast$ is the stellar luminosity, and $\sigma_\mathrm{SB}$ is the Stefan–Boltzmann constant. We compute this prior by varying $\phi$ between 0.02 and 0.05, typical of protoplanetary disks (e.g., \citealp{dsharp2}). We assume a Gaussian distribution for $L_\ast$, centered at 6.03 $L_\odot$, 2.57 $L_\odot$, and 0.36 $L_\odot$ for Elias 2-24, IM Lup, and DM Tau, respectively, following \citealp{andrews2018} and \citealp{jun2021}. The corresponding 1$\sigma$ distributions are 2.88 $L_\odot$, 1.23 $L_\odot$, and 0.18 $L_\odot$, as reported in the same references. Varying $T_\mathrm{d}$ allows us to introduce the intrinsic uncertainty of Eq. \ref{eqn:prior} into the posterior distributions of $\Sigma_\mathrm{d}$, $a_\mathrm{max}$, and $q$. Uniform priors are adopted for the remaining parameters. 

We explored the parameter space using the Markov chain Monte Carlo (MCMC) method implemented through the \texttt{emcee} package \citep{emcee}. The model parameters are varied within the following ranges: $T_\mathrm{d}$ between 3 and 100 K, $\Sigma_\mathrm{d}$ between $10^{-4}$ and 30 g cm$^{-2}$, $a_\mathrm{max}$ between $10^{-3}$ and $10^{2}$ cm (in logarithmic space), and $q$ between 1 and 4.5. We first run the fit with 32 walkers for 5000 steps, following a burn-in phase of 1000 steps. Across all radii and disks, the integrated autocorrelation times typically lie in the range of 60 - 80 steps, indicating that the number of steps satisfies standard convergence criteria. To reduce the impact of low probability multi-modal solutions, we perform a second run initialized around the median values from the first run, extending for 5000 steps after an additional burn-in of 1000 steps.  

As described in Sect. \ref{2.3}, all images were smoothed to a common beam size to minimize the effect of resolution differences on our dust characterization. Our model fitting is performed on the azimuthally averaged intensity profiles derived from these smoothed images, starting from $r = 11$ au to exclude the innermost regions potentially contaminated by free–free emission, and extending outward to the largest radius where emission is detected in the least sensitive observation i.e. 150, 300, and 300 au for Elias 2-24, IM Lup, and DM Tau, respectively (see Sect. \ref{3} for the definition of the halo's outer boundary). The profiles are sampled in radial steps of $n$, where $n$ corresponds to one-third of the spatial resolution of the smoothed images. The model is not convolved with the beam, since each radius is fitted independently. Thus, the effect of beam dilution depends on the final resolution of the smoothed images, becoming more significant at lower resolutions where emission is averaged over larger spatial scales. However, in the halo, where the intensity varies over spatial scales larger than the beam, the impact of beam dilution is less than in the inner disk. 

We adopted a Bayesian framework to infer the posterior probability distributions of these four parameters at each radius. The likelihood function is assumed to be Gaussian and can be expressed in log as:

\begin{equation}
\ln P(\bar{I}(r) | \Theta) = 
- \frac{1}{2} 
\sum_\mathrm{i}
\left[
\left(
\frac{\bar{I}_\mathrm{i} - I_{\mathrm{m},\mathrm{i}}}{\hat{\sigma}_{\bar{\mathrm{I}},\mathrm{i}}}
\right)^2
+ 
\ln\left(2\pi \hat{\sigma}_{\bar{\mathrm{I}},\mathrm{i}}^2\right)
\right],
\end{equation}

where $\Theta$ represents the vector of model parameters ($T_\mathrm{d}$, $\Sigma_\mathrm{d}$, $a_\mathrm{max}$, and $q$), $\bar{I}_\mathrm{i}$ is the observed azimuthally averaged intensity at radius $r$ and frequency $\nu_\mathrm{i}$, $I_{\mathrm{m},\mathrm{i}}$ is the model intensity at the same radius and frequency, and $\hat{\sigma}_{\bar{\mathrm{I}},\mathrm{i}}$ is the total uncertainty at that radius, defined as

\begin{equation}
\hat{\sigma}_{\bar{\mathrm{I}},\mathrm{i}} = 
\sqrt{
\sigma_{\bar{\mathrm{I}},\mathrm{i}}^2 + 
(\delta_\mathrm{i} \bar{I}_\mathrm{i})^2
}.
\end{equation}

Here $\delta_\mathrm{i}$ is the absolute flux calibration uncertainty at each frequency and, $\sigma_{\bar{\mathrm{I}},\mathrm{i}}$ denotes the mean error derived from the azimuthally averaged profiles (defined in Sect. \ref{3}). We assume flux calibration uncertainties of 20\% for Bands 9 and 8, 10\% for 7 and 6, and 5\% for Band 3. 

Furthermore, for Elias 2-24 and DM Tau, we adopt the double-resolution analysis technique introduced by \citealp{viscardi2025}. Essentially, we perform a second SED fit using the high resolution observations available: 0.87 mm, 1.3 mm, and 3 mm for Elias 2-24, and 0.9 mm and 1.3 mm for DM Tau, while using the $a_\mathrm{max}$ constraints derived from the mid–resolution fit (which incorporates all wavelengths) as a prior to guide this higher resolution analysis. 

\begin{figure*} 
 
    \includegraphics[width=1.0\textwidth]{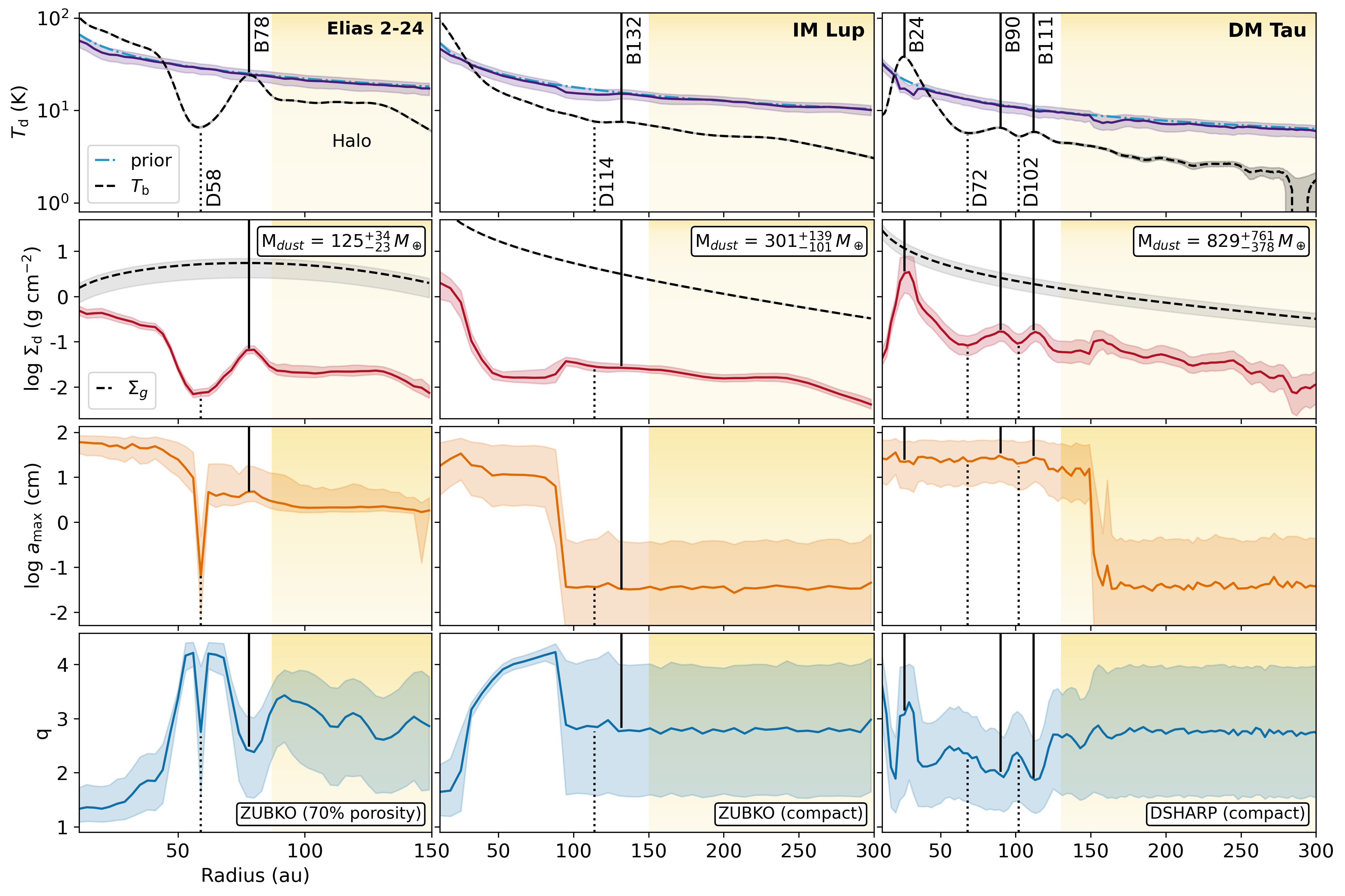} 

    \caption {\textit{First row:} Dust temperature, $T_\mathrm{d}$ (in purple), overplotted with the prior (Eq. \ref{eqn:prior}; dash-dotted blue line) and 1.3 mm brightness temperature, $T_\mathrm{b}$, in black dashes. The shaded regions represent the 1$\sigma$ uncertainty. For Elias 2-24 and DM Tau, the higher resolution $T_\mathrm{b}$ is plotted (8 $\times$ 8 au and 9 $\times$ 9 au, respectively) since the double resolution analysis was used to obtain these results. For IM Lup, the low resolution $T_\mathrm{b}$ (21 $\times$ 21 au) is plotted. Vertical solid and dotted lines indicate the rings and gaps, respectively. Yellow shaded bands mark the extent of the halo. All radial profiles start at 11 au. \textit{Second row:} Dust surface density, $\Sigma_\mathrm{d}$, along with the total mass of the disk calculated by extrapolating $\Sigma_\mathrm{d}$ down to the disk center and integrating over the disk. The gas surface density profiles, $\Sigma_\mathrm{g}$, defined in Sect. \ref{4.2}, are over-plotted in black dashes. \textit{Third row:} Maximum grain size, $a_\mathrm{max}$. \textit{Fourth row:} Grain size distribution slope, $q$. The preferred dust composition for each disk is indicated in the bottom-right corner of each panel.} \label{fig:results_main}
\end{figure*}

\begin{figure*} 
 
    \includegraphics[width=1.0\textwidth]{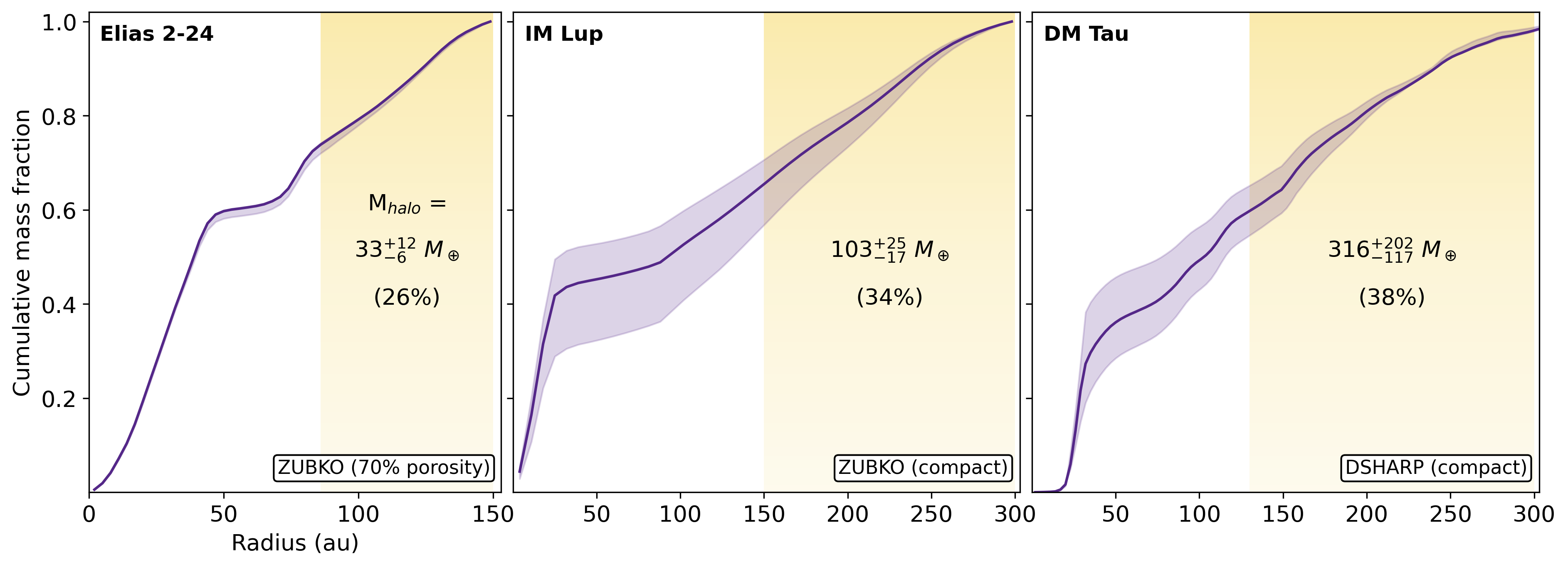} 
   \caption {{Cumulative mass fraction profiles are constructed by integrating $\Sigma_\mathrm{d}$ and summing the mass enclosed within successive annuli. The shaded regions represent the 1$\sigma$ uncertainty propagated from the posterior distributions. Yellow shaded bands mark the radial extent of the halo, with the corresponding mass and the percentage of the total disk mass it occupies. The preferred dust composition for each disk is indicated in the bottom-right corner of each panel.}} \label{fig:halo_mass}
\end{figure*} 

\subsection{SED modeling results}

Once we had the fits for all four dust compositions, we computed the Bayesian evidence of each MCMC fit with 500 random samples from the posterior. The Bayesian evidence, $Z$, for a model is given by $Z = \int L(D \mid \theta)\,\pi(\theta)\,d\theta$, where $L$ is the likelihood of the data with parameters $\theta$ and $\pi$ is the prior distribution. Then, we calculated the Bayes factor, $K$ with $K = {Z_1}/{Z_0}$ by adopting the DSHARP (compact) model as our reference (i.e., $Z_0$) (Fig. \ref{fig:bayes_factor}). We favor the Bayes factor over a traditional $\chi^{2}$ comparison because it naturally incorporates the influence of the prior. Following \citealp{guidi2022}, we referred to the scale proposed by \citealp{jeffreys1939} to interpret these results. We then identified the model with the highest number of radial points showing strong evidence as the most plausible description of the data. Below, we discuss the results of our fits to $T_\mathrm{d}$, $\Sigma_\mathrm{d}$, $a_\mathrm{max}$, and $q$ for the preferred composition, and their radial profiles are shown in Fig. \ref{fig:results_main}. In Appendix \ref{appendix_other_results}, we discuss in detail the results from the other compositions and their profiles are shown in Fig. \ref{fig:results_elias24}, \ref{fig:results_imlup}, and \ref{fig:results_dmtau}. To calculate the total dust mass of the disk, we extrapolated $\Sigma_\mathrm{d}$ inward from 11 au down to the disk center using the same radial spacing as the rest of the grid, assuming a constant surface density equal to the value at 11 au. Furthermore, we estimated the mass reservoir in the halo by integrating the dust surface density over the radial range corresponding to the region of smooth, featureless continuum emission (Fig. \ref{fig:halo_mass}).

\subsubsection{Elias 2-24}

The SED fitting for this disk is particularly challenging because of the sharp discontinuity in spectral index between the 1.3 – 3 mm and 0.87 – 1.3 mm profiles. In particular, some fits tend to converge toward unrealistically low temperatures, resulting in a high $\Sigma_\mathrm{d}$ and producing a massive disk (similar to \citealp{macias2021}). To mitigate this, we introduced two modifications to all our fits. First, we penalize $T_\mathrm{d}$ solutions that fall below the 5th percentile or above the 95th percentile of the Gaussian prior defined by Eq. \ref{eqn:prior}. Second, we include optically thicker mid–resolution observations at 0.74 mm (ALMA Band 8), allowing it to better constrain the midplane temperature. As discussed above, we utilize the double resolution analysis technique incorporating all four observations (smoothed to a common beam of 24 $\times$ 24 au), whose solution acts a prior for $a_\mathrm{max}$ for a higher resolution fit at 8 $\times$ 8 au, excluding the 0.74 mm observations. 

\begin{figure*} 
    
    \includegraphics[width=1.0\textwidth]{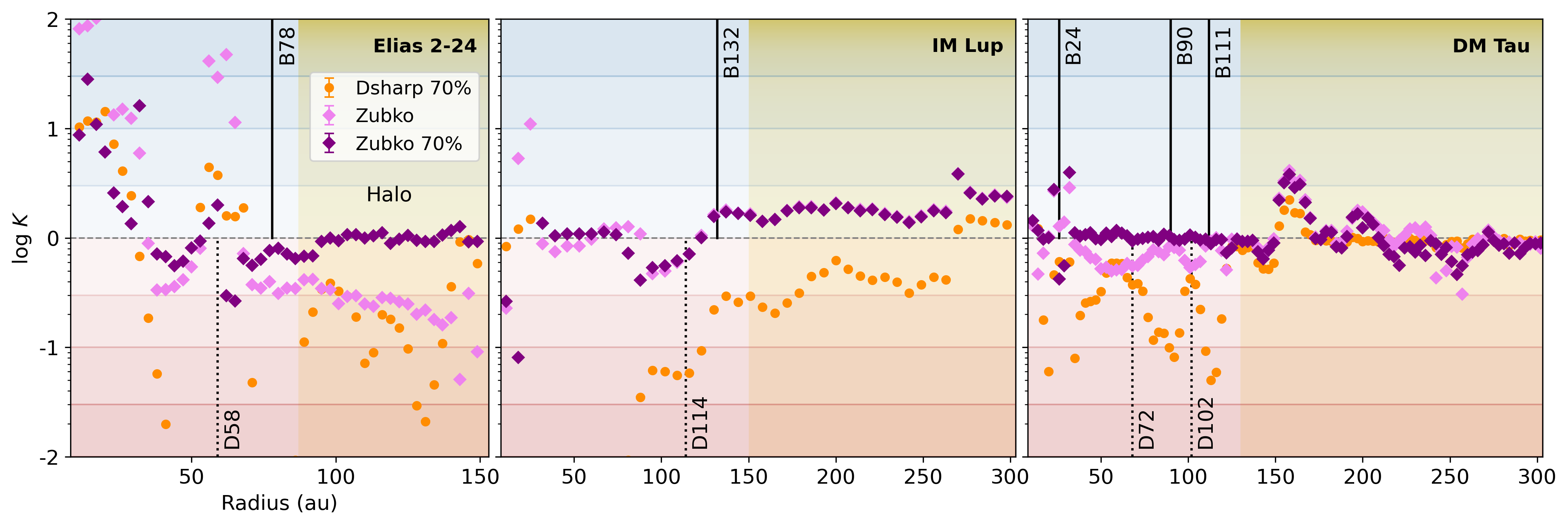} 
    \caption {{Bayes factor, $K$, for the DSHARP (70\% porosity), ZUBKO (compact), and ZUBKO (70\% porosity) models, computed relative to the DSHARP (compact) reference model. The blue color gradient indicates increasing evidence in favor of the corresponding model, while the red gradient indicates increasing evidence in favor of the reference model. Vertical solid and dotted lines indicate the rings and gaps, respectively. Yellow shaded bands mark the extent of the halo.}} \label{fig:bayes_factor}
\end{figure*}

To compare our results statistically, we then compute the Bayes factor ($K$) with DSHARP (compact) as our reference model. For the DSHARP (70\% porosity) model, 17\% of radii fall into the decisive–against category ($K < 0.01$), with an additional 40\% showing moderate to strong evidence against it ($0.01 < K < 0.3$). Only 11\% of radii provide anecdotal support ($1 < K < 3$), and a few fall into the moderate or strong categories, indicating that this model is systematically disfavored across most of the disk. For the ZUBKO (compact) model, 22\% of radii exhibit strong to decisive support ($K > 10$), although a large fraction (70\%) weakly or moderately disfavor it ($0.03 < K < 1$). For the ZUBKO (70\% porosity) model, 34\% of radii show anecdotal to moderate support ($1 < K < 10$), and 6\% show strong support ($K > 10$), while 55\% mildly disfavor it ($0.3 < K < 1$). Overall, we adopt the ZUBKO (70\% porosity) composition, as it is favored over the ZUBKO (compact) model in the halo (r > 96 au) (see Fig. \ref{fig:bayes_factor}), while also yielding temperature and grain size distributions that remain physically consistent with expectations from the $T_\mathrm{d}$ prior and radial drift theory, respectively. 

With the ZUBKO (70\% porosity) model, we find $\Sigma_\mathrm{d}$ = 0.2 - 0.45 g cm$^{-2}$ and $a_\mathrm{max}$ $\sim$ 50 cm at r < 40 au. Our model reproduces the local minima and maxima in $\Sigma_\mathrm{d}$ very well at D58 and B78, with values of 0.007 g cm$^{-2}$ and 0.06 g cm$^{-2}$, respectively, while the $a_\mathrm{max}$ at these locations is 1 mm and 4 cm, respectively. Compared to \citealp{carvalho2024} (who used DSHARP (compact)), at B78 our model finds $\Sigma_\mathrm{d}$ about a factor of 6 lower and $a_\mathrm{max}$ about a factor of 20 higher. In the halo (r > 96 au), $\Sigma_\mathrm{d}$ is mostly constant at 0.02 g cm$^{-2}$ until 123 au, and then it gradually decreases to 0.007 g cm$^{-2}$ at 150 au, with $a_\mathrm{max}$ mostly consistent at 2 cm. $q$ shows an overall increasing radial trend, rising from 1.3 to 4.2 at r < 56 au, after which it reaches a local minima of 2.7 at D58, followed by another local minima of 2.4 at B78. In the halo, $q$ remains approximately constant at 3.0, with uncertainties of 0.8. Following this, we extrapolate $\Sigma_\mathrm{d}$ down to 0 au and integrate over the disk, resulting in a total dust mass of 125$^{+34}_{-23}$ M$_{\oplus}$, a factor of 3 higher than the dust mass calculated by \citealp{carvalho2024}. Within the halo (r > 96 au), this corresponds to a dust mass of 33$^{+12}_{-6}$ M$_{\oplus}$ i.e. 26\% of the disk's total dust mass.

\subsubsection{IM Lup}

The 1.3 – 3 mm and 0.86 – 1.3 mm spectral index profiles are consistent with each other (bottom row of Fig. \ref{fig:profiles}), making the SED fit more manageable. Since high resolution data at two or more wavelengths are unavailable, this fit is performed at a smoothed resolution of 21 $\times$ 21 au, and beam dilution likely contributes to the suppressed temperatures. 

We then compare our results statistically by computing the Bayes factor ($K$). The DSHARP (70\% porosity) model is strongly disfavored, with 63\% of radii showing mild to strong evidence against it ($0.01 < K < 1$), 19\% showing decisively strong evidence against it ($K < 0.01$), and only 17\% showing anecdotal evidence ($1 < K < 3$). In contrast, the ZUBKO (compact) model is consistently favored across the disk: 70\% of radii fall in the anecdotal evidence range ($1 < K < 3$), with an additional 6\% showing moderate to strong evidence ($3< K < 30$), and only 20\% weakly disfavoring it ($0.3 < K < 1$). The ZUBKO (70\% porosity) model performs comparably well, with 76\% anecdotal evidence ($1 < K < 3$), though none show strong evidence, and 14\% of the radii weakly disfavor it. Overall, both ZUBKO compositions (compact and porous) are equally favored within the halo. Nevertheless, ZUBKO (compact) shows a slight statistical preference at some radii in the inner disk (where ZUBKO (70\% porosity) overestimates $\Sigma_\mathrm{d}$), making it the most robust choice for the analysis that follows. 

With the ZUBKO (compact) model, $\Sigma_\mathrm{d}$ roughly follows a power law decline at r < 40 au, decreasing from 2 g cm$^{-2}$ to 0.02 g cm$^{-2}$. At 40 < r < 250 au, $\Sigma_\mathrm{d}$ remains constant at $\sim$ 0.01 g cm$^{-2}$, and smoothly tapers off to 0.004 g cm$^{-2}$ by 300 au. At r < 90 au, $a_\mathrm{max}$ is $\sim$ 10 cm with an uncertainty of 0.7 dex, whereas at larger radii $a_\mathrm{max}$ remains mostly unconstrained with median values at 400 $\mu$m and uncertainties of 1 dex. The grain size distribution slope, $q$, exhibits a hyperbolic radial trend increasing from 1.6 to 4.2 at 90 au. Similar to $a_\mathrm{max}$, $q$ remains constant at 2.8 at r > 90 au with uncertainties of 1.2. Upon extrapolating $\Sigma_\mathrm{d}$ down to 0 au and integrating it, we find a total dust mass of 301$^{+139}_{-101}$ M$_{\oplus}$. Compared to \citealp{sierra2021}, who used DSHARP (compact), our model predicts $\Sigma_\mathrm{d}$ roughly an order of magnitude lower, resulting in a disk four times less massive, and $a_\mathrm{max}$ one to two orders of magnitude higher at r < 90 au. In comparison to \citealp{francheschi2023}, who modeled the dust distribution using VLT/SPHERE and 1.3 mm ALMA observations, our disk mass is also lower by a factor of four. Within the halo (r > 150 au), our $\Sigma_\mathrm{d}$ corresponds to a dust mass of 103$^{+25}_{-17}$ M$_{\oplus}$, i.e., 34\% of the disk's total dust mass. 

\subsubsection{DM Tau}

The SED fitting of this disk is particularly interesting because it exhibits optically thick emission between 0.45 – 0.9 mm and transitions to optically thin behavior between 0.9 – 1.3 mm (bottom row of Fig. \ref{fig:profiles}). Due to the availability of high resolution observations at least at two wavelengths, we utilize the double resolution analysis technique to map the changes in $\Sigma_\mathrm{d}$ more precisely: we perform a mid resolution fit at a common beam size of 40 $\times$ 40 au incorporating all four observations, followed by a high resolution fit at 9 $\times$ 9 au with the 0.9 and 1.3 mm observations where the solution from the former acts as a prior for $a_\mathrm{max}$. Furthermore, we find that the high optical depth of B24 (Fig. \ref{fig:optical_depth}) leads to an overestimation of $\Sigma_\mathrm{d}$. Therefore, we refit a few radii around B24 with a reduced parameter space for $\Sigma_\mathrm{d}$: $10^{-4}$ to 10 g cm$^{-2}$. 

\begin{figure*} 
    \includegraphics[width=1.0\textwidth]{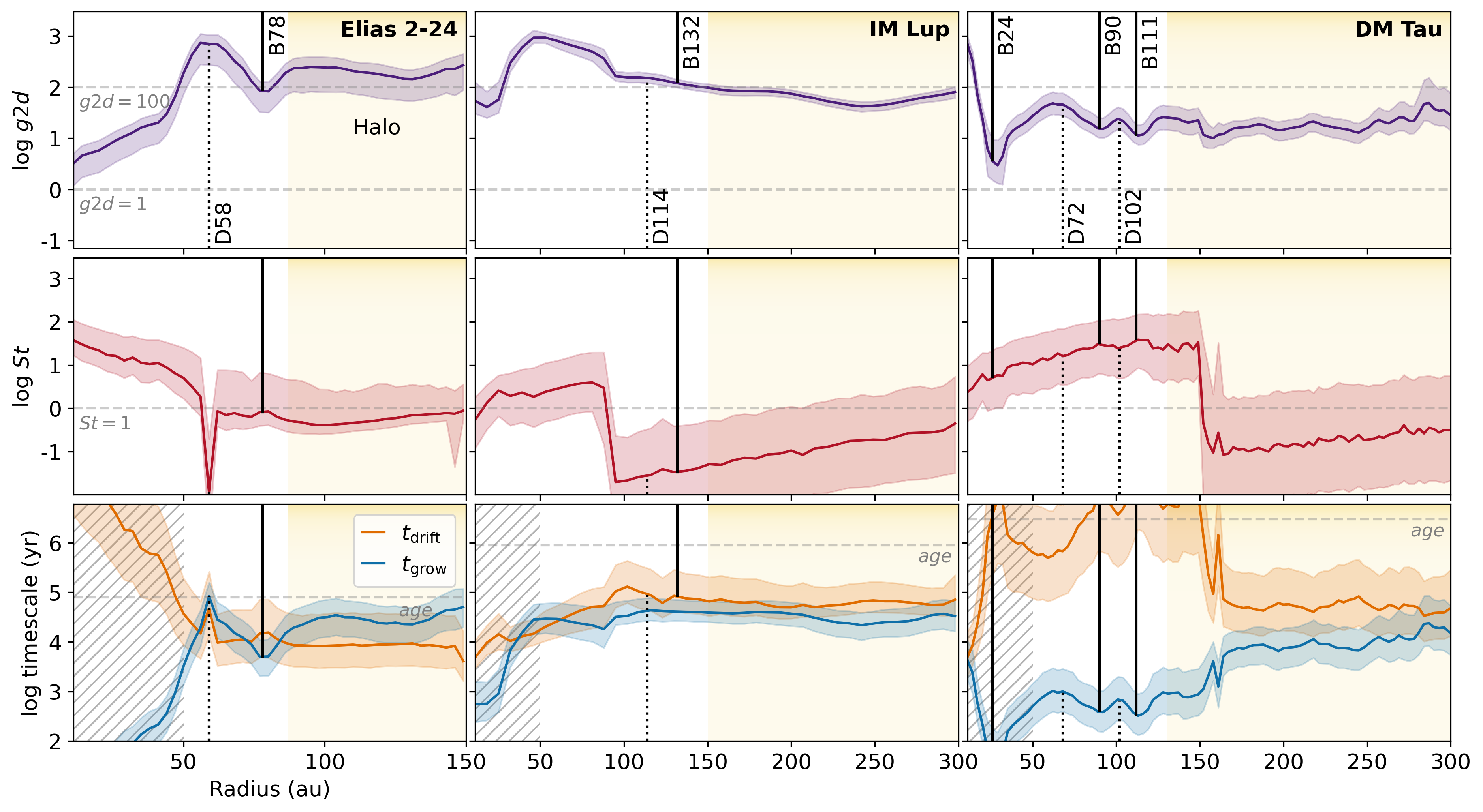} 
    \caption {\textit{Top row:} Gas-to-dust ratio, $g2d =\Sigma_\mathrm{g}/\Sigma_\mathrm{d}$. Gray horizontal dashed lines indicate $g2d$ = 1 and $g2d$ = 100. Vertical solid and dotted lines indicate the rings and gaps, respectively. Yellow shaded bands mark the extent of the halo. \textit{Middle row:} Stokes number, St, given by Eq. \ref{eqn:stokes}. The gray horizontal dashed line indicates St = 1. \textit{Bottom row:} Drift, $t_\mathrm{drift}$, and growth timescales, $t_\mathrm{grow}$, given by Eq. \ref{eqn:drift} and Eq. \ref{eqn:grow}, respectively. The horizontal dashed line indicates the lower limit of the disk's age. The inner 50 au are indicated with gray hatching, as these regions are affected by both beam dilution and higher optical depth, rendering the inferred timescales less robust.}\label{fig:timescales}
\end{figure*}

The Bayes factor ($K$) distributions indicate that the DSHARP (70\% porosity) model is strongly disfavored across most of the disk: about 67\% of radii fall into the weak-against category ($0.3 < K < 1$), with an additional 21\% showing moderate to strong evidence against ($0.03 < K < 0.3$), while only 12\% provide anecdotal support ($1 < K < 3$) and none show stronger positive evidence. The ZUBKO (compact) model performs better, with 32\% of radii showing anecdotal to moderate support ($1 < K < 10$), although a majority (68\%) still weakly disfavor it. The ZUBKO (70\% porosity) model shows the highest fraction of radii with positive support among the other models, with 35\% exhibiting anecdotal to moderate evidence ($1 < K < 10$), but it remains weakly disfavored at nearly two-thirds of the radii. It appears that ZUBKO (70\% porosity) is favored in the inner disk, while the DSHARP (compact) offers the best description of the halo (r > 130 au). Overall, none of the alternative compositions achieve strong or decisive evidence over the reference, and the predominance of weak-against classifications for all three confirms DSHARP (compact) as the most statistically favored description of the data. 

With the DSHARP (compact) model, we find that $\Sigma_\mathrm{d}$ rises steeply from 0.09 to 3 g cm$^{-2}$ from the disk center to B24, where it reaches a local maxima, and then declines sharply, exhibiting successive local minima and maxima of 0.08, 0.17, 0.09, and 0.16 g cm$^{-2}$ at D72, B90, D102, and B111, respectively. In the halo (r > 130 au), $\Sigma_\mathrm{d}$ decreases more gradually, from 0.06 to 0.008 g cm$^{-2}$. For r < 130 au, $a_\mathrm{max}$ has median values of $\sim$ 25 cm with uncertainties of 0.5 dex, while at larger radii $a_\mathrm{max}$ becomes largely unconstrained, with median values of 800 $\mu$m and uncertainties of 1 dex. The inferred $q$ values show clear radial structure, with median values of 3.5, 2.3, 1.9, 2.4, and 1.9 at B24, D72, B90, D102, and B111, respectively, and uncertainties of 0.5. In the halo (r > 130 au), $q$ is unconstrained, with median values of 2.7 and uncertainties of 1.2. Extrapolating $\Sigma_\mathrm{d}$ down to 0 au and integrating over the disk yields a total dust mass of 829$^{+761}_{-378}$ M$_{\oplus}$. Our $\Sigma_\mathrm{d}$ and $a_\mathrm{max}$ results are consistent with \citealp{sierra2024} who also used DSHARP (compact) to model the disk in the inner 50 au. Within the halo (r > 130 au), we infer a dust mass of 316$^{+202}_{-117}$ M$_{\oplus}$, corresponding to 38\% of the disk's total dust mass. 

\subsection{Evolutionary timescales} \label{4.2}

Following the SED modeling, we calculate $g2d$ (the gas-to-dust ratio) given by $\Sigma_\mathrm{g}/\Sigma_\mathrm{d}$, where $\Sigma_\mathrm{g}$ is the gas surface density. For Elias 2-24, we assume a total gas-to-dust mass ratio of 100 and adopt the scale radius $R_\mathrm{c}$ and power law index from \citealp{pinte2023}. For IM Lup and DM Tau, we utilize the $\Sigma_\mathrm{g}$ derived from dynamical gas mass estimates by \citealp{martire2024} (with vertical stratification) and \citealp{longarini2025}, respectively. Therefore, the disk gas masses used in this work for Elias 2-24, IM Lup, and DM Tau are 0.038$^{+0.009}_{-0.007}$, 0.106$^{+0.002}_{-0.002}$, and 0.057$^{+0.019}_{-0.020}$ M$_{\odot}$, respectively. The $\Sigma_\mathrm{g}$ profiles are shown in the second row of Fig. \ref{fig:results_main}. We also estimated the Stokes number as a function of $a_\mathrm{max}$ given by, 
\begin{equation} \label{eqn:stokes}
\mathrm{St} = \frac{\pi \, \rho_{\mathrm{m}} \, a_\mathrm{max}}{2 \, \Sigma_\mathrm{g}},
\end{equation}
where $\rho_{\mathrm{m}} = (1 - \mathcal{P}) \left( \sum_\mathrm{i} \frac{f_\mathrm{i}}{\rho_\mathrm{i}} \right)^{-1}$ is the dust bulk density. Here $f_\mathrm{i}$ are the mass fractions of the individual components, $\rho_\mathrm{i}$ their material densities, and $\mathcal{P}$ is the porosity. For DSHARP (compact), DSHARP (70\% porosity), ZUBKO (compact), and ZUBKO (70\% porosity), $\rho_{{m}}$ is 1.675, 0.503, 2.036, and 0.611 g cm$^{-3}$, respectively. St quantifies the degree of coupling between dust and gas \citep{whipple}: St >> 1 indicates weak coupling to the gas, whereas St << 1 implies that the dust is strongly coupled. Furthermore, we calculated $t_\mathrm{drift}$ (drift timescale) and $t_\mathrm{grow}$ (growth timescale) as \citealp{birnstiel2024}, 
\begin{equation} \label{eqn:drift}
t_{\mathrm{drift}} =
\left(\frac{r}{h_\mathrm{g}}\right)^{2}
\left( \mathrm{St} + \mathrm{St}^{-1} \right)
(1+\epsilon)^2
\frac{1}{\gamma \Omega_\mathrm{K}},
\end{equation}

\begin{equation} \label{eqn:grow}
t_{\mathrm{grow}} =
\begin{cases}
\displaystyle
\sqrt{\frac{\alpha_\mathrm{t}}{\alpha_\mathrm{t} + \mathrm{St}}}
\left(\frac{r}{h_\mathrm{g}}\right)
\frac{g2d}{\Omega_\mathrm{K}},
& \mathrm{St} > \dfrac{\alpha_\mathrm{t} r}{h_\mathrm{g}} \\[10pt]

\displaystyle
\frac{2}{\sqrt{3}}
\frac{g2d}{\Omega_\mathrm{K}},
& \mathrm{St} \le \dfrac{\alpha_\mathrm{t} r}{h_\mathrm{g}}.
\end{cases}
\end{equation}

The first regime (St $> \alpha_\mathrm{t} r / h_\mathrm{g}$) applies when the drift velocity dominates the relative velocity, while the second regime (St $\le \alpha_\mathrm{t} r / h_\mathrm{g}$) applies when turbulent velocity dominates. In the above equations, $h_\mathrm{g}$ = $c_\mathrm{s}/\Omega_\mathrm{K}$ is the gas scale height assuming hydrostatic equilibrium, and $\Omega_\mathrm{K} = \sqrt{\frac{G\,M_{\star}}{r^{3}}}$ is the Keplerian angular frequency at radius $r$, gravitational constant $G$, and stellar mass $M_{\star}$. The isothermal sound speed is $c_\mathrm{s} = \sqrt{\frac{k_\mathrm{B} T_\mathrm{d}}{\mu m_\mathrm{H}}}$, where $k_\mathrm{B}$ is the Boltzmann constant, $\mu$ is the mean molecular weight, and $m_\mathrm{H}$ is the hydrogen mass. The dimensionless pressure gradient parameter is defined as $\gamma = \left| \frac{d \ln P}{d \ln r} \right|$, where $P$ is the gas pressure. Assuming an isothermal vertical structure, $P = \Sigma_\mathrm{g}c_\mathrm{s}\Omega_\mathrm{K}$. The midplane dust-to-gas volume density ratio is $\epsilon = \frac{\Sigma_\mathrm{d}}{\Sigma_\mathrm{g}} \sqrt{1 + \frac{\mathrm{St}}{\alpha_\mathrm{t}}}$, where $\alpha_\mathrm{t}$ is the turbulent viscosity parameter. 

For Elias 2-24, IM Lup and DM Tau, we adopt $M_{\star}$ of 0.78$^{+0.34}_{-0.13}$, 1.19$^{+0.002}_{-0.002}$, and 0.47$^{+0.014}_{-0.015}$ M$_{\odot}$, respectively \citep{andrews2018, martire2024, longarini2025}. To propagate uncertainties into the timescales, we perform bootstrapping with 5000 samples at each radius, drawing $\Sigma_\mathrm{d}$, $\Sigma_\mathrm{g}$, $T_\mathrm{d}$, St, and $M_{\star}$ from gaussian distributions defined by their lower and upper bounds. Estimates of $\alpha_\mathrm{t}$ for Elias 2-24 can be found in \citealp{carvalho2024} and \citealp{villenave2025}, IM Lup in \citealp{francheschi2023} and \citealp{flaherty2024}, and DM Tau in \citealp{flaherty2020} and \citealp{careno2024}. Due to the wide range of these estimates, we adopt a conservative range of $10^{-4} \le \alpha_\mathrm{t} \le 10^{-2}$ that we sample uniformly. The median and 1$\sigma$ spread of the resulting distribution at each radius is taken as the nominal value and associated uncertainty of $t_\mathrm{drift}$ and $t_\mathrm{grow}$. The resulting radial profiles are given in Fig. \ref{fig:timescales} and we discuss below their features in the halo. 

\paragraph{Elias 2-24.} The resulting profile exhibits a $g2d \sim 200$ in the halo. Furthermore, we find St $\sim 1$, indicating marginally decoupled dust grains. We obtain $t_\mathrm{drift} = 8 \times 10^{3}$ yr and $t_\mathrm{grow} = 3 \times 10^{4}$ yr, both 1-2 orders of magnitude shorter than the estimated disk age of 0.08 - 0.5 Myr \citep{andrews2018}. 

\paragraph{IM Lup.} We find a $g2d \sim 64$ in the halo. Due to the larger uncertainties in $a_\mathrm{max}$, the inferred St carries significant uncertainties; however, throughout most of the halo we find St $\leq 1$, consistent with marginal dust–gas coupling. The corresponding timescales are $t_\mathrm{drift} = 6 \times 10^{4}$ yr and $t_\mathrm{grow} = 3 \times 10^{4}$ yr, both approximately 1-2 orders of magnitude shorter than the estimated disk age of 0.9 - 1.3 Myr \citep{avenhaus2018}. 

\paragraph{DM Tau.} In the halo, $g2d \sim$ 20. The larger uncertainties in $a_\mathrm{max}$ propagate into St, but we find $\mathrm{St} \leq 1$, suggesting marginal dust–gas coupling. The inferred timescales are $t_\mathrm{drift} = 8 \times 10^{4}$ yr and $t_\mathrm{grow} = 10^{4}$ yr, are again 1-2 orders of magnitude shorter than the estimated disk age of 2.9 - 6 Myr \citep{garufi2024}.

\begin{table*}
\caption{Properties of the halo.}
\label{table:results}
\centering
\begin{tabular}{c c c c c c c c c c}    
\hline\hline       
\thead{Disk} & \thead{Preferred \\ composition} & \thead{$T_\mathrm{d}$ (K)} &  \thead{$\Sigma_\mathrm{d}$ (g cm$^{-2}$)} & \thead{$a_\mathrm{max}$ (cm)} & \thead{$q$} & \thead{$g2d_\mathrm{mass}$} & \thead{log St} & \thead{log $t_\mathrm{drift}$ \\ (yr)} & \thead{log $t_\mathrm{grow}$ \\ (yr)} \\ [3ex]

\midrule
\vspace{1mm}
Elias 2-24 & ZUBKO (70\% porosity) & 19.7$^{+2.6}_{-3.1}$ & 0.019$^{+0.007}_{-0.004}$ & 2.13$^{+3.55}_{-0.63}$ & 3.0$^{+0.7}_{-1.1}$ & 195.9$^{+109.6}_{-127.8}$ & -0.2$^{+0.7}_{-0.2}$ & 3.9$^{+0.6}_{-0.4}$ & 4.5$^{+0.4}_{-0.4}$ \\ 
\vspace{1mm}
IM Lup & ZUBKO (compact) & 11.9$^{+1.4}_{-1.4}$ & 0.015$^{+0.003}_{-0.002}$ & 0.04$^{+0.34}_{-0.03}$ & 2.8$^{+1.2}_{-1.2}$ & 64.2$^{+14.5}_{-13.5}$ & -0.7$^{+1.0}_{-1.1}$ & 4.8$^{+0.4}_{-0.2}$ & 4.5$^{+0.2}_{-0.3}$\\ 
\vspace{1mm}
DM Tau & DSHARP (compact) & 7.2$^{+1.0}_{-1.2}$ & 0.042$^{+0.027}_{-0.016}$ & 0.08$^{+0.81}_{-0.08}$ & 2.7$^{+1.2}_{-1.2}$ & 17.5$^{+19.6}_{-10.5}$ & -0.7$^{+1.3}_{-1.2}$ & 4.9$^{+0.9}_{-0.7}$ & 4.0$^{+0.5}_{-0.4}$ \\ 

\hline\hline    
\vspace{0.1mm}
\end{tabular}
\newline Notes: The various properties are averaged over the entire radial extent of the halo, except $g2d_\mathrm{mass}$ which is the gas-to-dust mass ratio in the halo.
\end{table*}

\section{Discussion} \label{5} 

\subsection{Origin of halos - Evidence of late infall}

In IM Lup and DM Tau, the dust drift and growth timescales are shorter by 1–2 orders of magnitude than the lower limits on the disk ages. Additionally, we find $t_\mathrm{grow} \leq t_\mathrm{drift}$, indicating that grain growth should proceed efficiently before radial drift removes particles. If the dust in the halo was primordial and resulted solely from long-term viscous spreading of the disk (e.g., \citealp{trapman2020}), the short drift timescales would require the presence of pressure traps to prevent the rapid loss of solids (e.g., \citealp{dullemond2018}, \citealp{pinilla2025}). This favors an alternative scenario in which the halo material is not in steady state but instead reflects relatively recent delivery of dust, for example through late infall from the surrounding environment \citep{dullemond2019, winter2024}. In this picture, dust grains deposited at large radii (e.g., \citealp{kuffmeier2020}) would not yet have had sufficient time to grow and drift inward, naturally explaining the persistence of extended emission as well as the small grain sizes. While viscous spreading may still contribute to the overall disk extent, our results suggest that late-time mass loading could provide an explanation for the origin and survival of the smooth halos in IM Lup and DM Tau. 

Similarly, for Elias 2-24, the inferred timescales are 1-2 orders of magnitude shorter than the lower limit on the disk's age. Such a short drift timescale relative to the system age, together with the cm-sized grains, indicates that hidden substructures, perhaps unresolved and affected by a high optical depth, must be slowing radial drift and keeping these grains in the halo. In addition, the $^{12}$CO channel maps presented by \citealp{andrews2018} show signs of cloud contamination and an irregular gas morphology, which may support cloudlet capture. The combination of rapid growth and survival against drift therefore favors a role for hidden pressure traps, potentially also triggered by late infall (e.g., \citealp{zhao2025}). 

In this context, \citealp{winter2024} show that late infall from the ISM can supply sufficient material to sustain higher stellar accretion rates. Elias 2–24, IM Lup, and DM Tau have mass accretion rates of $6 \times 10^{-7}$, $10^{-8}$, and $10^{-8}$ M$_{\odot}\,\mathrm{yr}^{-1}$, respectively \citep{manara2023}. These accretion rates are consistent with a scenario in which Elias 2–24 may be experiencing ongoing infall, whereas IM Lup and DM Tau may have undergone infall earlier in their evolution.

\subsection{Occurrence rate of halos}

Recent ALMA surveys of nearby star-forming regions reach continuum sensitivities of 2 K\footnote{The brightness temperatures in this section were calculated with the full Planck function.} in Taurus at 1.3 mm \citep{long2018,long2019}, 2 – 8 K in Lupus at 1.3 mm \citep{alvarado2025}, 3 K in Upper Sco at 0.88 mm \citep{barenfeld2016}, and 2 K in Ophiuchus at 1.3 mm \citep{cieza2019}. At 1.3 mm, the halo emission in Elias 2-24, IM Lup, and DM Tau spans brightness temperatures of 14 – 6 K, 7 – 3 K, and 5 – 2 K at the inner and outer halo edges, respectively. While the halos of Elias 2-24 and IM Lup would therefore be detectable in most of these surveys, the halo of DM Tau could potentially be missed in the Lupus survey which has a median sensitivity of 4 K. This demonstrates that halo occurrence rates are biased by survey sensitivity. An example of a survey with sufficient sensitivity is the exoALMA large program, in which most disks exhibit halo-like extended emission \citep{teague2025}. However, it is not clear whether these structures share the same dust evolution pathways as those analyzed in this work. 

For example, \citealp{curone2025} show a clear correlation between disk size and the steepness of the outer continuum profile, with larger disks exhibiting flatter outer slopes. Additionally, the halo slopes are steeper at longer wavelengths, causing the outermost halo emission to taper off quickly and become increasingly difficult to detect (Fig. \ref{fig:profiles}). Moreover, outer disks with steeper slopes may reflect a different evolutionary regime than the flatter halos discussed in this work. A possible explanation for steeper slopes is that differential radial drift increases with radius and with particle size, effectively truncating the outer dust disk and creating a sharp outer edge. In contrast, disks such as IM Lup and DM Tau with higher local dust-to-gas ratios or slower drift relative to growth (Fig. \ref{fig:timescales}), maintain a more extended population of solids at large radii, resulting in a shallower outer slope \citep{birnstiel2014}. The inferred halo masses constitute a substantial fraction of the total dust mass, contributing up to 25 – 40\%. While halos alone do not fully resolve the mass budget problem \citep{manara2018}, they represent a non-negligible reservoir of material that is systematically missed in many surveys. Assuming that micron-sized grains emit efficiently at sub-mm wavelengths, one way to assess the presence of a halo is to obtain deep short baseline observations (i.e. at lower spatial resolution) with sufficient LAS (largest angular scale) and compare them to long baseline data to quantify how much flux is not recovered in the higher resolution images. 

\subsection{Extent of the dust mass in large disks}

Our results indicate that $R_{90}$ does not necessarily enclose 90\% of the total dust mass, particularly in the presence of extended halos where a substantial mass reservoir could exist with little contribution to the total flux. In our sources, the halo alone accounts for 20 – 30\% of the emission, implying that $R_{90}$ captures only part of this extended component. Since the outermost 10\% of flux can be distributed over large spatial scales, moving from $R_{90}$ to $R_{95}$ or $R_{99}$ can significantly increase the disk size. For example, in IM Lup and DM Tau, including an additional 9\% of the flux (i.e. from $R_{90}$ to $R_{99}$) at 1.2 mm and 0.9 mm, respectively, expands the disk size by $\sim$ 100 au (Table \ref{table:disk_sizes}). This suggests that higher flux fractions provide a more representative proxy for the spatial extent of the dust mass, consistent with the findings of \citealp{rosotti2019}. However, these metrics also depend on wavelength. The dust mass radius can be more reliably constrained at shorter wavelengths (e.g., at 0.87 mm; ALMA Band 7) where up to 40\% of the dust mass may reside in sub-mm grains (see IM Lup and DM Tau in Fig. \ref{fig:halo_mass}), whereas at longer wavelengths the lower opacity of small grains reduces the detectability of the outer disk. 

\subsection{Gas-to-dust ratios in IM Lup and DM Tau}

The halos of IM Lup and DM Tau exhibit gas-to-dust mass ratios of 64 and 18, respectively, which are below the ISM value of 100. However, both systems have disk-to-star mass ratios exceeding 5\%, for which the gas mass can be robustly inferred through dynamical mass measurements to within 25\% accuracy \citep{veronesi2024}, lending confidence to these estimates. At the same time, the $^{12}$CO gas disks extend to radii of $\sim$ 1000 au in both sources \citep{cleeves2016, galloway2025}, potentially due to late infall, whereas the $R_{99}$ dust disks sizes (Table \ref{table:disk_sizes}) are smaller by a factor of 3. This suggests substantial radial drift, with dust grains migrating inward while the gas disk remains extended. In this picture, the low gas-to-dust ratios in the halos may reflect dust evolution, whereas beyond the dust edge the gas-to-dust ratio is likely higher than 100.

\subsection{Turbulence in Elias 2-24}

The flat radial profile of $a_\mathrm{max}$ (Fig. \ref{fig:results_main}) in the halo suggests that the grains there may be fragmentation limited. Following \citealp{jiang2024}, we can estimate the level of turbulence with $\alpha_\mathrm{frag} = v_\mathrm{frag}^2/3\mathrm{St}c_\mathrm{s}^2$. While \citealp{jiang2024} predict grains to be fragmentation limited at $v_\mathrm{frag} =$  1 ms$^{-1}$, we consider a broader range of 1 - 10 ms$^{-1}$. Using the $T_\mathrm{d}$ and St values from Fig. \ref{fig:results_main} and Fig. \ref{fig:timescales}, respectively, we obtain $10^{-6} < \alpha_\mathrm{frag} < 2\times10^{-3}$ in the halo. Higher turbulent viscosity enhances vertical stirring, keeping a larger fraction of grains suspended in the disk atmosphere and preventing them from settling at the midplane. Therefore, if $\alpha_\mathrm{frag}$ lies near the upper bound, the outer disk is expected to appear more flared in near-infrared emission (e.g., with VLT/SPHERE) than in the case of lower turbulence levels. 

\section{Conclusions} \label{6} 

We identify faint, extended emission in three of the brightest protoplanetary disks: Elias 2–24, IM Lup, and DM Tau, which we define as ``Halos''. We define the halo as the outer disk region where the radial intensity profile transitions into a smooth, featureless continuum. For Elias 2–24, IM Lup, and DM Tau, the halo extends beyond radii of 86, 150, and 130 au, respectively, and the halos in our sources account for 20 - 30\% of the total flux density at ALMA wavelengths. Spectral index profiles reveal markedly different behaviors in the halo across the three systems. In Elias 2–24, the disk is optically thick between 0.74 – 0.87 mm and 0.87 – 1.3 mm, with $\alpha \sim 2$ throughout the halo, and becomes optically thin between 1.3 – 3 mm, where $\alpha \sim 3$. The halo of IM Lup is characterized by intermediate values of $2 < \alpha < 3$ between 0.86 - 1.3 mm and 1.3 - 3 mm, indicative of partially optically thin emission. In contrast, the halo of DM Tau exhibits $1 < \alpha < 2$ at 0.45 – 0.9 mm and $2 < \alpha < 4$ at 0.9 – 1.3 mm, a behavior that can only be explained when scattering is taken into account \citep{sierra2020}.

To better understand the physical nature and origin of the halo, we model the spectral energy distribution (SED) using ALMA continuum observations at four (sub)millimeter wavelengths. We carry out all fits using four dust compositions: DSHARP (compact), DSHARP (70\% porosity), ZUBKO (compact), and ZUBKO (70\% porosity). By evaluating the Bayes factors relative to DSHARP (compact), we identify the preferred compositions for Elias 2-24, IM Lup, and DM Tau as ZUBKO (70\% porosity), ZUBKO (compact), and DSHARP (compact), respectively. Using the preferred composition for each disk, we find that the halo of Elias 2–24 consists of large cm-sized grains, while the halos of IM Lup and DM Tau consist of smaller grains, with $a_\mathrm{max} < 4$ mm and $a_\mathrm{max} < 9$ mm, respectively. The total dust masses of Elias 2-24, IM Lup, and DM Tau are $125^{+34}_{-23}$, $301^{+139}_{-101}$, and $829^{+761}_{-378}$ M$_{\oplus}$, with corresponding halo masses of $33^{+12}_{-6}$, $103^{+25}_{-17}$, and $316^{+202}_{-117}$ M$_{\oplus}$. As a result, halos make up 25 - 40\% of the total dust content, demonstrating that they represent a substantial reservoir of solids and might play an important role in alleviating the mass-budget problem. 

Finally, we combine our dust constraints with gas surface density profiles from the literature. We find $g2d_\mathrm{mass}$ = 196, 64, and 18 in the halos of Elias 2-24, IM Lup and DM Tau, respectively. The halos of IM Lup and DM Tau are dust rich, which may reflect ongoing grain growth and radial drift in a recently replenished outer disk, which may also explain their $\sim$ 1000 au large gas disks. Moreover, with the Stokes numbers we find marginally to fully coupled grains in the halos of all three sources.

Our key result is that dust growth and drift timescales in the halos are 1–2 orders of magnitude shorter than the respective disk ages, indicating that the smooth outer disks should not exist. In IM Lup and DM Tau, we also find $t_\mathrm{grow} \leq t_\mathrm{drift}$, implying that grain growth should proceed efficiently before radial drift removes particles. The survival of their halos therefore favors a scenario in which the outer disks are replenished by relatively recent material delivery through late infall. This explains both the small grain sizes and why the dust has not yet had sufficient time to drift inward. Elias 2-24's halo also exhibits short timescales relative to its age, but together with cm-sized grains this suggests the presence of pressure traps, in the form of unresolved substructures and possibly affected by a high optical depth, that allowed grain growth. Intriguingly, the $^{12}$CO channel maps in \citealp{andrews2018} suggest that late infall may still contribute to the formation of the halo. We also estimate that if the grains in the halo of Elias 2-24 are fragmentation limited with $v_\mathrm{frag} =$ 1 - 10 ms$^{-1}$, we can have turbulence of $10^{-6} < \alpha_\mathrm{frag} < 2\times10^{-3}$. 

\section*{Data availability}
The data associated with this paper is publicly available and can be found in \url{https://zenodo.org/records/20051905}. \\ 

\begin{acknowledgements}
We thank the referee for an insightful report. This paper makes use of the following ALMA data: 2013.1.00498.S, 2016.1.00484.L, 2016.1.00565.S, 2017.1.01330.S, 2017.1.01460.S, 2018.1.01055.L, 2018.1.01119.S, 2018.1.01198.S, 2018.1.01755.S, 2019.1.00837.S, 2019.1.01357.S, 2019.1.01760.S, 2021.1.00378.S, 2021.1.00879.S, 2021.1.01123.L. ALMA is a partnership of ESO (representing its member states), NSF (USA) and NINS (Japan), together with NRC (Canada), MOST and ASIAA (Taiwan), and KASI (Republic of Korea), in cooperation with the Republic of Chile. The Joint ALMA Observatory is operated by ESO, AUI/NRAO and NAOJ. SD and TB acknowledge funding from the Deutscher Akademischer Austauschdienst (DAAD; grant number 57693453), the European Union under the European Union’s Horizon Europe Research and Innovation Programme 101124282 (EARLYBIRD) and funding by the Deutsche Forschungsgemeinschaft (DFG, German Research Foundation) under Germany’s Excellence Strategy - EXC-2094 - 390783311. Views and opinions expressed are, however, those of the authors only and do not necessarily reflect those of the European Union or the European Research Council. Neither the European Union nor the granting authority can be held responsible for them. PC acknowledges support from the Italian Ministero dell’Istruzione, Università e Ricerca through the grant Progetti Premiali 2012 – iALMA (CUP C52I13000140001), and from the ANID BASAL project FB210003. 
\end{acknowledgements}

\bibliographystyle{aa} 
\bibliography{ref}

\clearpage
\begin{appendix}

\section{DM Tau at 0.45 mm}\label{appendix_dmtau}

\begin{figure}[H]
    \includegraphics[width=0.5\textwidth]{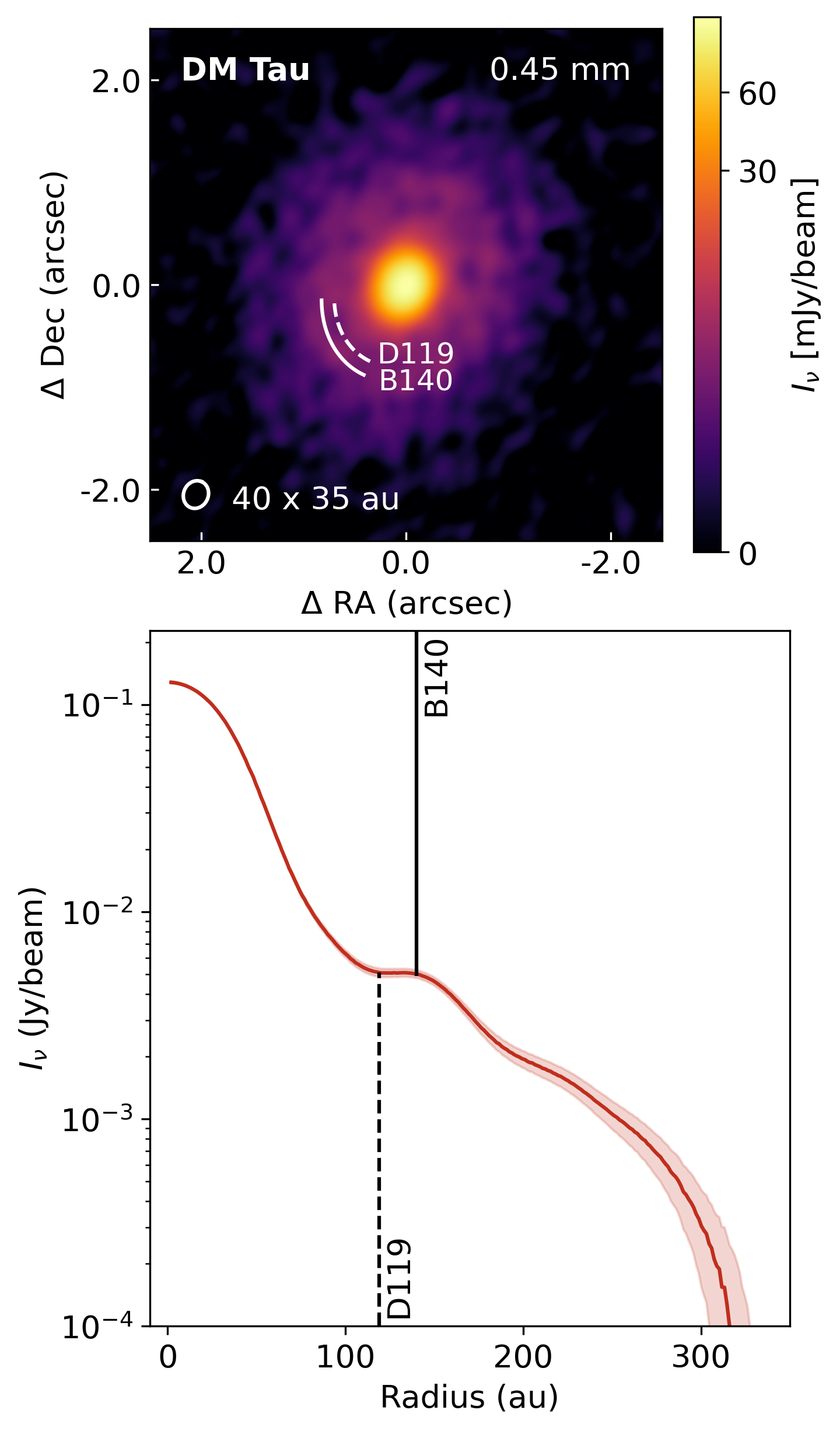} 
    \caption {Continuum image and intensity profile at 0.45 mm.}\label{fig:dmtau_B9}
\end{figure}

In Fig. \ref{fig:dmtau_B9}, we highlight the newly identified gap D119 and ring B140 revealed by the ALMA Band 9 observations at 0.45 mm. As these substructures are absent in the higher-resolution intensity profiles at 0.9 mm and 1.3 mm, they are likely produced by optical depth effects. In particular, the 0.45 mm emission traces higher vertical layers of the disk, where variations in temperature can imprint rings and gaps that are not present deeper in the midplane (see Fig. \ref{fig:optical_depth} where $\tau \sim 0.3$ near B140).

\section{Opacity and optical depth}\label{appendix_opacity}

\begin{figure}[H]
    \includegraphics[width=0.5\textwidth]{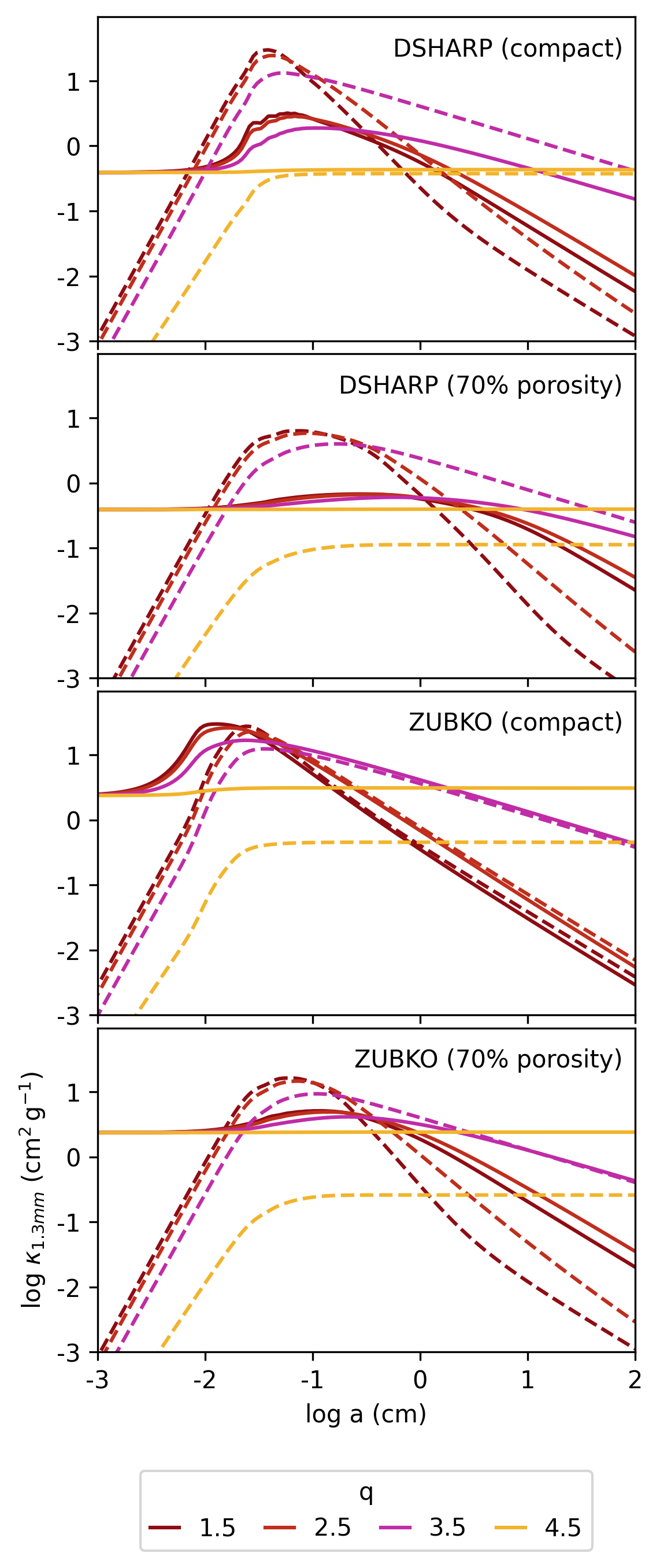} 
    \caption {Absorption (solid) and scattering (dashed) opacity laws, $\kappa$ of the four compositions at 1.3 mm for different grain size distribution slopes, $q$.}\label{fig:opacity_laws}
\end{figure}

\begin{figure*} 
    \includegraphics[width=1.0\textwidth]{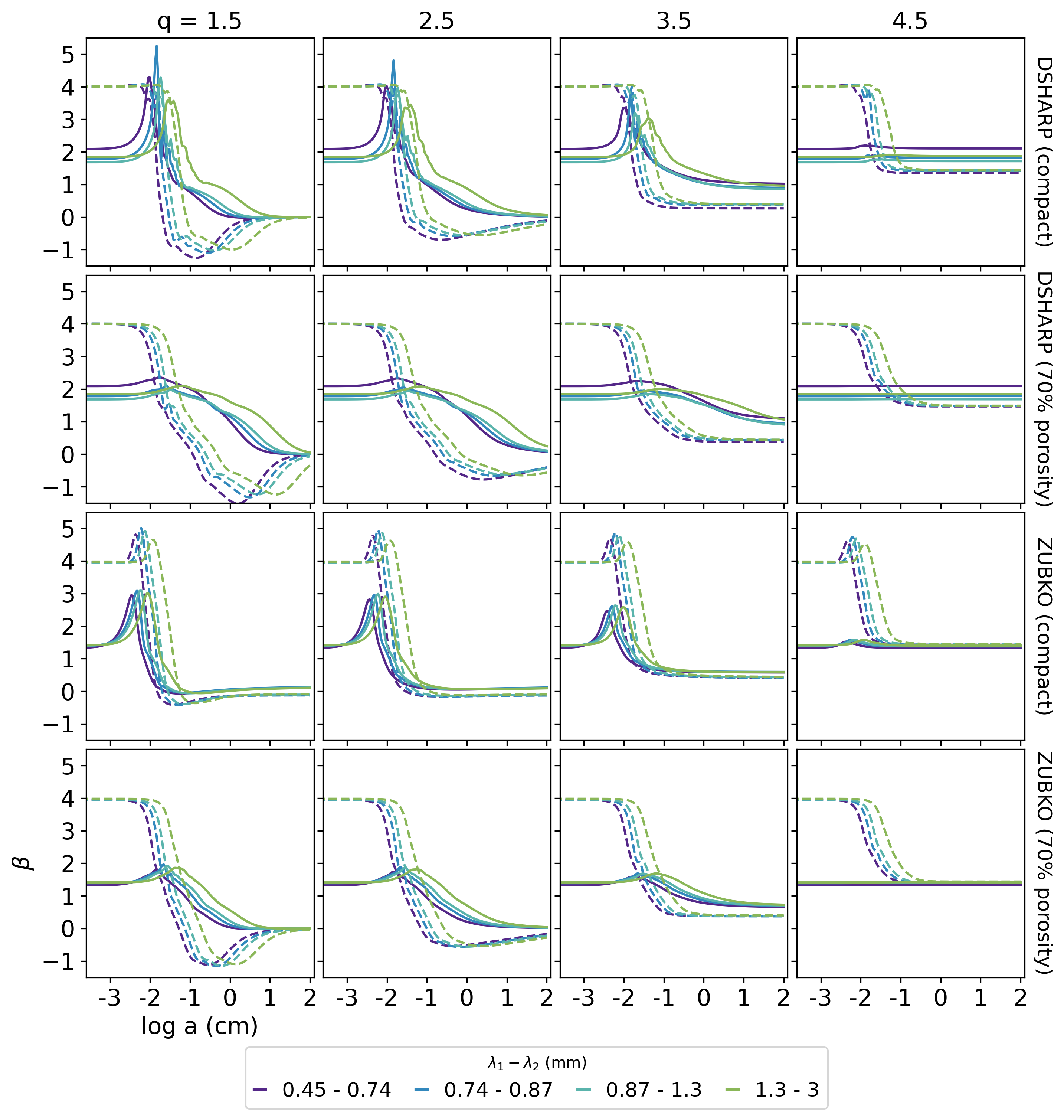} 
    \caption {Spectral index, $\beta$ of the absorption (solid) and scattering (dashed) opacity of the four compositions between five wavelengths for different grain size distribution slopes, $q$. Each row represents the composition listed on the right whereas each column represents the $q$ listed on top.}\label{fig:beta}
\end{figure*}

\begin{figure*}
    \includegraphics[width=1.0\textwidth]{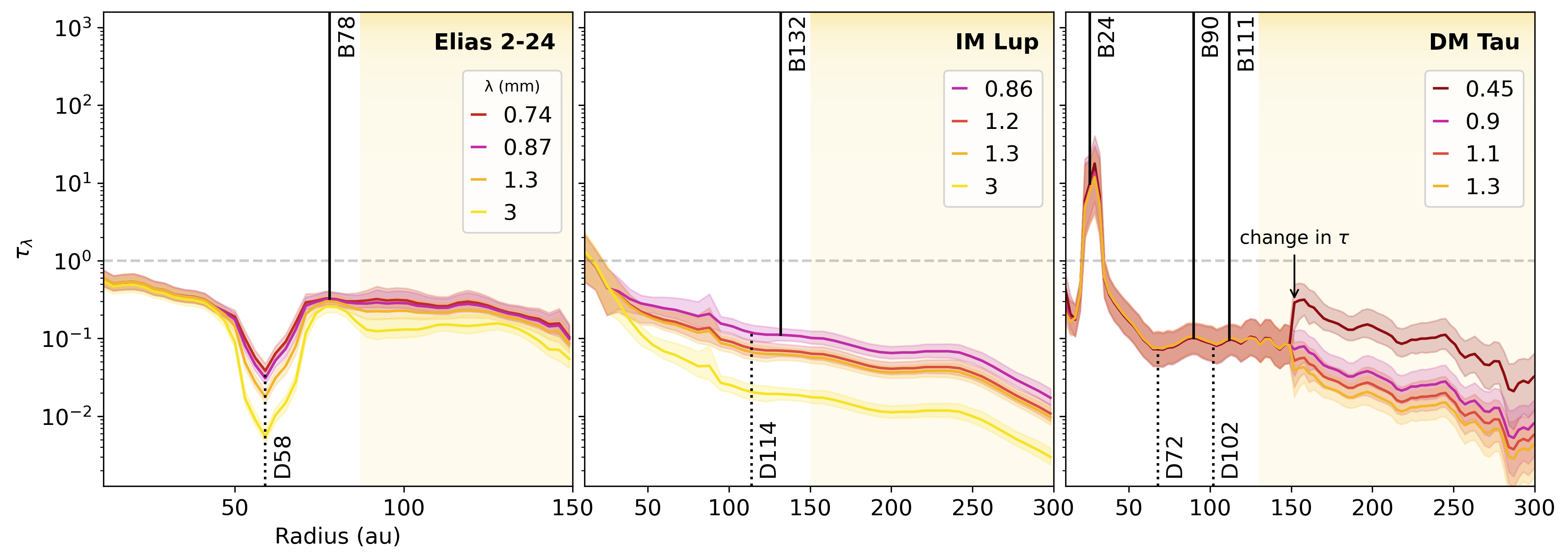} 
    \caption {Optical depth, $\tau_\lambda$, posterior distributions with the preferred compositions ($\Sigma_\mathrm{d}$, $a_\mathrm{max}$, and $q$ are taken from Fig. \ref{fig:results_main}). Vertical solid and dotted lines indicate the gaps and rings, respectively. In addition to the previously identified gaps and rings, we highlight the region near B140 at 0.45 mm in DM Tau where $\tau$ changes abruptly. They gray horizontal dashed line indicates $\tau = 1$. Yellow shaded bands mark the extent of the halo.}\label{fig:optical_depth}
\end{figure*}

Fig. \ref{fig:opacity_laws} shows the $\kappa$ (absorption and scattering opacity) laws of the four compositions used in the modeling of the multiwavelength observations of the three disks. Fig. \ref{fig:beta} shows $\beta$ (spectral index of the absorption and scattering opacity) between the five wavelengths used in this work. Finally, Fig. \ref{fig:optical_depth} shows the $\tau_\lambda$ (optical depth) posterior distributions for the fits with the preferred composition. 

\section{SED modeling results with the other compositions}\label{appendix_other_results}

\begin{figure*}
    \centering 
    \includegraphics[width=1.0\textwidth]{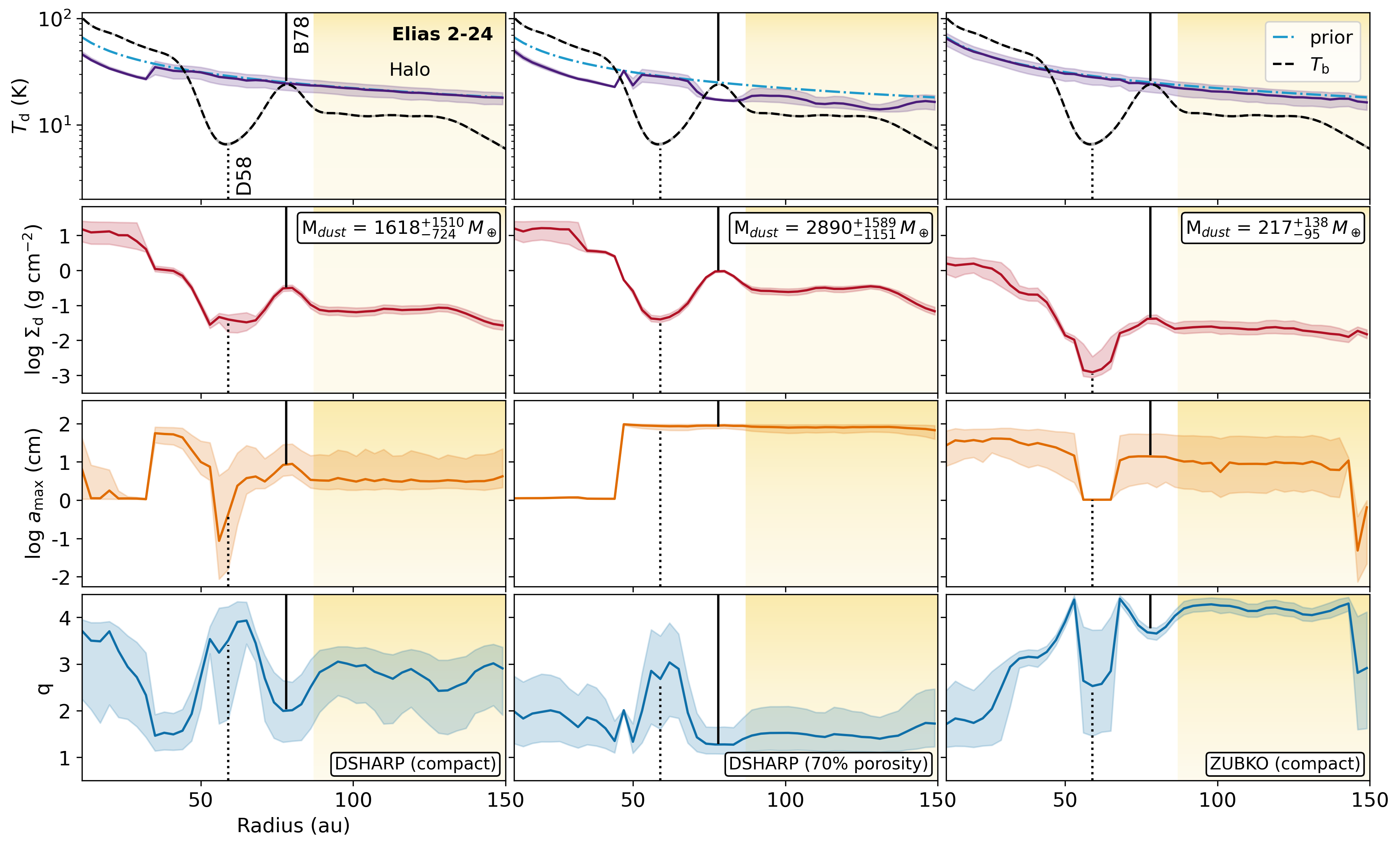} 
    \caption {Models of Elias 2-24 with the non-preferred compositions.} \label{fig:results_elias24}
\end{figure*}

\begin{figure*} 
    \includegraphics[width=1.0\textwidth]{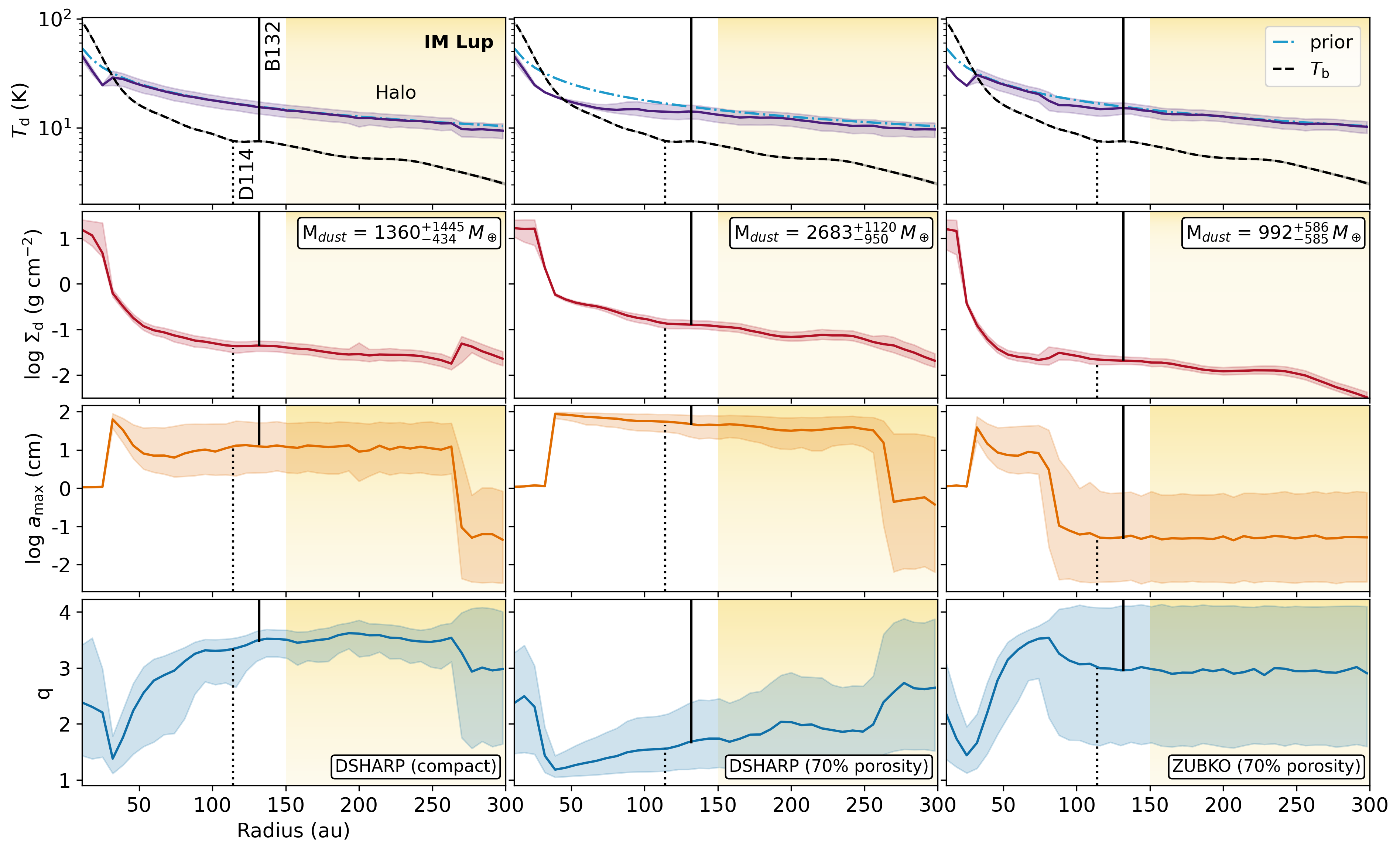} 
    \caption {Models of IM Lup with the non-preferred compositions.} \label{fig:results_imlup}
\end{figure*}

\begin{figure*} 
    \includegraphics[width=1.0\textwidth]{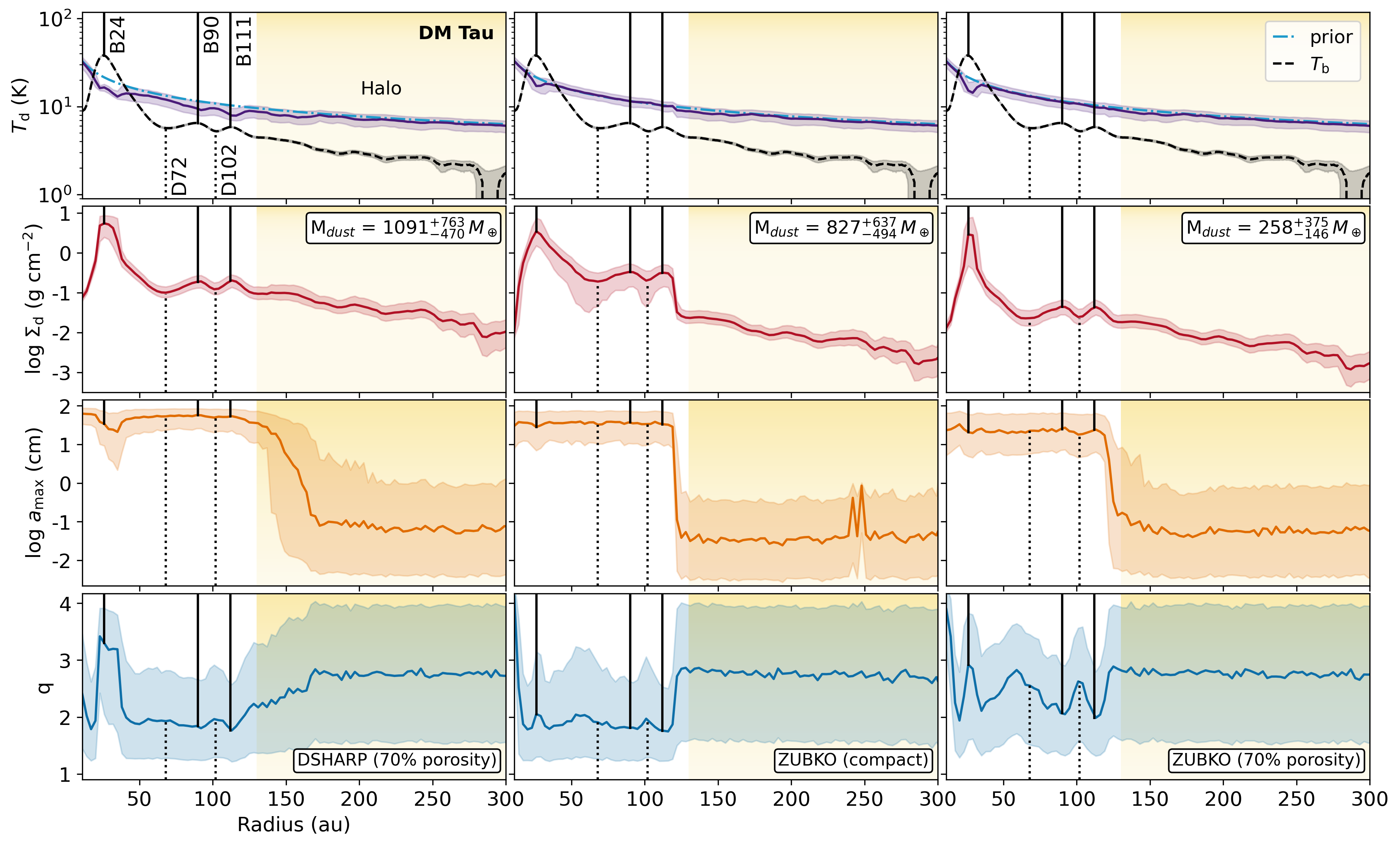} 
    \caption {Models of DM Tau with the non-preferred compositions.} \label{fig:results_dmtau}
\end{figure*}

Fig. \ref{fig:results_elias24}, \ref{fig:results_imlup}, \ref{fig:results_dmtau} present the SED modeling results for Elias 2-24, IM Lup and DM Tau, respectively, with the non-preferred compositions. Below, we compare these results in detail to those derived with the preferred composition shown in Fig. \ref{fig:results_main}.


\subsection{Elias 2-24}

Since we impose a strong prior on $T_\mathrm{d}$, the temperature solutions remain consistent across all four opacity models within 1$\sigma$. A notable trend is that both DSHARP models (compact and porous) yield temperatures that are $\sim$10 K lower than the values expected in the inner 50 au, similar to the findings of \citealp{macias2021} for the inner 20 au of TW Hya. This suggests that the ZUBKO models (both compact and porous) are better suited for reproducing the higher temperatures expected in the inner disk. For the dust surface density, the DSHARP (70\% porosity) model predicts $\Sigma_\mathrm{d}$ values that are roughly a factor of 3 higher than those from the DSHARP (compact) model, consistent with \citealp{guidi2022}, who found 5 times higher $\Sigma_\mathrm{d}$ values for HD 163296 with porous DSHARP grains. In contrast, the ZUBKO models (compact and porous) generally produce similar $\Sigma_\mathrm{d}$ values, except in the inner 35 au where ZUBKO (compact) is a factor of 5 higher, and at D58, where ZUBKO (70\% porosity) yields values about 7 times higher. Overall, the ZUBKO models give surface densities that are approximately 3 times lower than DSHARP (compact) and 12 times lower than DSHARP (70\% porosity). 

Regarding the maximum grain size, $a_\mathrm{max}$, the DSHARP (70\% porosity) model predicts grains about an order of magnitude larger than those from DSHARP (compact), driven by its preference for lower $T_\mathrm{d}$ and higher $\Sigma_\mathrm{d}$. The ZUBKO models agree with each other within uncertainties. A striking feature is that the DSHARP models infer smaller grains in the inner disk (r < 50 au) but predict grain sizes 10 – 70 times larger in the halo, inconsistent with expectations from radial drift, which should concentrate the largest grains toward the inner disk. In contrast, the ZUBKO models produce larger grains in the inner disk and grain sizes about 5 times smaller in the halo, aligning with expectations of the radial drift theory \citep{whipple}. For $q$, the DSHARP models exhibit a decreasing radial trend at r < 40 au, while the ZUBKO models show an increasing one. The DSHARP models reach a local minima at D58, in contrast to the ZUBKO models, which display a local maxima at the same location, and all four models converge to a local minimum at B78. In the halo (r > 96 au), DSHARP (compact) and ZUBKO (70\% porosity) yield mutually consistent $q$ values, whereas ZUBKO (compact) predicts values higher by a factor of 1.4 relative to these models and by a factor of 2.8 compared to DSHARP (70\% porosity).

\subsection{IM Lup}

The DSHARP (70\% porosity) model tends to settle at temperatures that are 5 - 10 K lower at r < 100 au, whereas the $T_\mathrm{d}$ prior suggests 20 – 45 K in this region. The DSHARP (compact) and ZUBKO (70\% porosity) models perform somewhat better, following $T_\mathrm{d}$ closely except the solution at r = 25 au which converges to $\sim$ 25 K instead of the expected 35 K. In contrast, the ZUBKO (compact) model follows the prior temperature profile very well throughout the disk. 

Similar to Elias 2–24, the DSHARP (70\% porosity) model yields $\Sigma_\mathrm{d}$ values that are higher than those from DSHARP (compact), by factors of two – four across most of the disk, with the contrast gradually diminishing at larger radii. The two ZUBKO models, in turn, remain consistent with each other within uncertainties, except at $r < 25$ au where ZUBKO (70\% porosity) produces median $\Sigma_\mathrm{d}$ values an order of magnitude higher than those from ZUBKO (compact), driven by its preference for lower $T_\mathrm{d}$. In general, the $\Sigma_\mathrm{d}$ values given by DSHARP (70\% porosity) are a factor of 7 - 20 higher and that given by DSHARP (compact) are a factor of two to four higher than the ones given by the ZUBKO models, with the disparity decreasing toward larger radii. 

In terms of $a_\mathrm{max}$, the DSHARP models are consistent with each other within uncertainties, but DSHARP (70\% porosity) has median values that are an order of magnitude higher than the values predicted by DSHARP (compact) at r < 250 au and a factor of eight higher at r > 270 au. This again reflects its bias toward lower $T_\mathrm{d}$ and higher $\Sigma_\mathrm{d}$. DSHARP's  compact model also predicts $q$ values about a factor of 2 higher than the porous one. The ZUBKO models agree with each other within uncertainties except at r < 25 au where the compact model predicts $a_\mathrm{max}$ that are an order of magnitude higher than the porous model. ZUBKO's compact model predicts $q$ that are a factor of 1.2 higher than the porous one at r < 90 au, while the two are consistent at larger radii. As in Elias 2–24, the DSHARP compositions predict cm-sized grains throughout most of the disk (r < 270 au), whereas the ZUBKO compositions infer grain sizes more in line with expectations from radial drift: larger grains concentrated at r < 90 au and smaller grains dominant at larger radii.

\subsection{DM Tau}

The $T_\mathrm{d}$ solutions remain consistent with the prior across all four opacity models. In terms of $\Sigma_\mathrm{d}$, the two DSHARP models yield mutually consistent surface densities, while the ZUBKO (compact) model produces values that are roughly an order of magnitude higher than those of ZUBKO (70\% porosity) at r < 120 au, converging to similar values at larger radii. Overall, ZUBKO (70\% porosity) predicts $\Sigma_\mathrm{d}$ values about a factor of 5 lower, and ZUBKO (compact) about a factor of 2 higher, than the DSHARP models. Importantly, all four models successfully reproduce the local minima and maxima associated with the gaps and rings. For $a_\mathrm{max}$, all models infer comparable trends that are consistent with the radial drift theory, with larger grains at r < 115 au and smaller, largely unconstrained grains at larger radii. The $q$ profiles also trace the disk substructure, with local maxima at B24, D72, and D102 and local minima at B90 and B111. At r > 115 au, $q$ remains mostly unconstrained. 

\onecolumn
\twocolumn[\section{Observation log}\label{appendix_obs}]
\begin{table}
\vspace{2mm}
\caption{\raggedright{Summary of the ALMA observations.}}
\label{table:data}
\centering
\begin{tabular}{c c c c c c c c c}     
\hline\hline       
                     
\thead{Band} & \thead{Project Code} & \thead{P.I.} &  \thead{Freq. (GHz)} & \thead{Mous ID \\ (uid://A001/..)} & \thead{Observation \\ Date} & \thead{Baselines (m)} & \thead{On-source \\ time (min)} & \thead{CASA \\ version \tablefootmark{a}} \\ [3ex]

\midrule
\multicolumn{9}{c}{\textbf{Elias 2-24}} \\
\midrule

3 & 2017.1.01330.S & L. Perez & 89.5 - 105.5 & X1284/X117a & 2017 Nov 30 & 15 - 2500 & 14.92 & 5.4.0 \\ 
& 2018.1.01198.S & L. Perez & 89.5 - 105.5 & X1359/X31 & 2019 Jun 12 & 244 - 16200 & 34.37 & 5.4.0 \\
& & & & & 2019 Jun 14 & 244 - 16200 & 26.61 & 5.4.0 \\
& & & & & 2019 Jun 20 & 244 - 16200 & 33.57 & 5.4.0 \\
& & & & & 2019 Jun 22 & 244 - 16200 & 32.05 & 5.4.0 \\
& 2019.1.01760.S & R. Booth & 89.6 - 105.4 & X1465/X1dd & 2021 Jul 12 & 28 - 3396 & 30.14 & 6.2.1 \\  
& & & & X1465/X1db & 2021 Sep 09 & 178 - 16196 & 33.77 & 6.2.1 \\
& & & & & 2021 Sep 11 & 178 - 16196 & 31.85 & 6.2.1 \\

\midrule

6 & 2013.1.00498.S & L. Perez & 216.0 - 233.4 & X13a/Xed & 2015 Jul 21 & 15 - 14950 & 11.09 & 4.3.1 \\
& 2016.1.00484.L & S. Andrews & 230.1 - 247.9 & X8c5/X4e & 2017 Sep 25 & 15 - 14950 & 28.59 & 4.7.2 \\
&  &  &  & & 2017 Oct 03 & 15 - 14950 & 29.74 & 4.7.2 \\
&  &  &  & & 2017 Oct 04 & 15 - 14950 & 29.74 & 4.7.2 \\

\midrule

7 & 2021.1.00879.S & L. Perez & 343.5 - 358.0 & X159f/Xa9 & 2022 Jun 12  & 15 - 1213 & 37.60 & 6.2.1 \\
&  &  & & & 2022 Jul 01 & 15 - 1301 & 37.60 & 6.2.1 \\
&  &  & & X159f/Xa7 & 2023 Jun 27 & 92 - 8548 & 42.99 & 6.2.1 \\
&  &  & & & 2023 Jul 01 & 85 - 8548 & 42.30 & 6.2.1 \\
&  &  & & & 2023 Jul 01 & 85 - 8548 & 43.04 & 6.2.1 \\
&  &  & & & 2023 Jul 03 & 85 - 8548 & 43.28 & 6.2.1 \\

\midrule

8 & 2021.1.00378.S & L. Cieza & 397.1 - 413.0 & X1590/X2734 & 2022 Aug 04 & 15 - 1302 & 0.77 & 6.2.1 \\
& & & & & 2022 Aug 11 & 15 - 1302 & 0.87 & 6.2.1 \\

\midrule 
\multicolumn{9}{c}{\textbf{IM Lup}} \\
\midrule

3 & 2017.1.01330.S & L. Perez & 89.5 - 105.5 & X1284/X1186 & 2017 Nov 30 & 113 - 13894 & 21.87 & 5.1.1 \\ 
& & & & & 2017 Nov 30 & 113 - 13894 & 21.87 & 5.1.1 \\
& 2018.1.01055.L & K. Oberg & 86.0 - 101.4 & X133d/X19a8 & 2018 Oct 29 & 15 - 1398 & 36.45 & 5.4.0 \\
& & & & X133f/X35 & 2018 Nov 06 & 15 - 1398 & 79.49 & 5.4.0 \\
& & & & X133f/X33 & 2019 Aug 22 & 41 - 3638 & 35.78 & 5.4.0 \\
& & & & & 2019 Aug 22 & 41 - 3638 & 35.78 & 5.4.0 \\

\midrule

6 & 2013.1.00226.S & K. Oberg & 216.1 - 235.2 & X122/X1e9 & 2014 Jul 06 & 20 – 650 & 20.68 & 4.2.2 \\
& & & 241.5 - 260.5 & X122/X1fd & 2014 Jul 17 & 20 – 650 & 21.21 & 4.2.2 \\
& 2013.1.00694.S & I. Cleeves & 241.0 - 260.3 & X121/X2cb & 2015 Jan 29 & 15 – 349 & 17.10 & 4.2.2 \\
& & & 241.0 - 260.3 & X121/X2c9 & 2015 May 13 & 21 – 558 & 33.79 & 4.2.2 \\
& 2013.1.00798.S & C. Pinte & 216.9 - 233.7 & X122/X5de & 2015 Jun 10 & 21 – 784 & 37.37 & 4.2.2 \\
& 2016.1.00484.L & S. Andrews & 230.1 - 247.9 & X8c5/X72 & 2017 Sep 25 & 41 - 14851 & 30.49 & 5.1.1 \\
& & & & & 2017 Oct 24 & 41 - 13894 & 30.48 & 5.1.1 \\ 
& 2018.1.01055.L & K. Oberg & 217.2 - 234.9 & X133d/X19d2 & 2018 Nov 28  & 15 - 1241 & 41.93 & 5.4.0 \\
& & & & X133d/X19d0 & 2019 Aug 11  & 41 - 3638 & 84.37 & 5.4.0 \\
& & & & & 2019 Aug 12  & 41 - 3638 & 114.27 & 5.4.0 \\
& & & & & 2019 Aug 14  & 41 - 3638 & 82.86 & 5.4.0 \\
& & & 248.1 - 266.0 & X133d/X19e7 & 2019 Apr 07  & 15 - 500 & 16.01 & 5.4.0 \\
& & & & & 2019 Apr 09  & 15 - 500 & 64.09 & 5.4.0 \\
& & & & X133d/X19e5 & 2019 Aug 20  & 41 - 3397 & 47.41 & 5.4.0 \\
& & & & & 2019 Aug 20  & 41 - 3397 & 47.31 & 5.4.0 \\ 

\hline\hline    
\end{tabular}
\vspace{2mm}
\newline \raggedright{Notes: \tablefoottext{a}{CASA version used for data calibration.}}
\end{table}

\begin{table*}
\caption{\raggedright{Table \ref{table:data} continued.}}
\centering
\begin{tabular}{c c c c c c c c c}    
\hline\hline       
                       
\thead{Band} & \thead{Project Code} & \thead{P.I.} &  \thead{Freq. (GHz)} & \thead{Mous ID \\ (uid://A001/..)} & \thead{Observation \\ Date} & \thead{Baselines (m)} & \thead{On-source \\ time (min)} & \thead{CASA \\ version \tablefootmark{a}} \\ [3ex]

\midrule
\multicolumn{9}{c}{\textbf{IM Lup}} \\
\midrule

7 & 2019.1.01357.S & R. Teague & 339.4 - 351.5 & X1465/Xf17 & 2019 Nov 26 & 15 - 314 & 45.23 & 5.6.1 \\
& & & & & 2019 Nov 27 & 15 - 314 & 45.33 & 5.6.1 \\
& & & & X1465/Xf15 & 2019 Oct 06 & 15 - 1547 & 11.09 & 6.1.1 \\
& & & & & 2021 Apr 01 & 15 - 1040 & 47.17 & 6.1.1 \\
& & & & & 2021 Apr 03 & 15 - 1397 & 47.07 & 6.1.1 \\
& & & & & 2021 Apr 03 & 15 - 1263 & 47.17 & 6.1.1 \\
& & & & & 2021 Apr 08 & 15 - 1040 & 47.17 & 6.1.1 \\

\midrule
\multicolumn{9}{c}{\textbf{DM Tau}} \\
\midrule

6 & 2013.1.00498.S & L. Perez & 215.9 - 233.4 & X13a/Xe5 & 2015 Aug 12 & 15 - 1574 & 14.20 & 6.6.1 \\
& 2017.1.01460.S & J. Hashimoto & 216.1 - 233.6 & X1284/X230b & 2017 Oct 27 & 135 - 14851 & 28.63 & 6.6.1 \\
& & & & & 2017 Oct 27 & 135 - 14851 & 28.63 & 6.6.1 \\
& 2018.1.01755.S & J. Hashimoto & 212.5 - 230.9 & X133d/X4079 & 2019 Jun 05 & 83 - 15238 & 39.51 & 6.6.1 \\
& & & & & 2019 Jun 05 & 83 - 15238 & 39.51 & 6.6.1 \\
& & & & & 2019 Jun 05 & 83 - 15238 & 39.51 & 6.6.1 \\

\midrule

7 & 2018.1.01119.S & K. Flaherty & 277.0 - 291.0 & X15a1/Xe33 & 2018 Aug 25 & 15 - 1398 & 46.85 & 5.4.0 \\
& & & & & 2018 Nov 04 & 15 - 1398 & 46.85 & 5.4.0 \\
& & & & & 2018 Nov 04 & 15 - 1398 & 46.85 & 5.4.0 \\
& & & & & 2018 Nov 05 & 15 - 1398 & 46.85 & 5.4.0 \\
& & & & & 2018 Nov 05 & 15 - 1398 & 46.85 & 5.4.0 \\
& 2021.1.01123.L & R. Teague & 330.6 - 345.8 & X133d/X2c5e & 2018 Aug 25 & 9 - 2696 & 331.65 & 5.4.0 \\

\midrule

9 & 2016.1.00565.S & K. Schwarz & 658.2 - 676.7 & X87c/X44f & 2018 Aug 16 & 15 - 479 & 40.32 & 6.6.1 \\
& 2019.1.00837.S & K. Flaherty & 670.0 - 691.6 & X1465/X220e & 2019 Oct 29 & 15 - 697 & 45.76 & 6.6.1 \\
& & & & & 2022 Sep 22 & 15 - 500 & 45.76 & 6.6.1 \\

\hline\hline    
\vspace{1mm}
\end{tabular}

\end{table*}
\end{appendix}

\end{document}